\documentclass[fleqn,usenatbib]{mnras}

\usepackage{newtxtext,newtxmath}

\usepackage[T1]{fontenc}

\DeclareRobustCommand{\VAN}[3]{#2}
\let\VANthebibliography\thebibliography
\def\thebibliography{\DeclareRobustCommand{\VAN}[3]{##3}\VANthebibliography}

\usepackage[dvipdfmx]{graphicx}	
\usepackage{amsmath}	
\usepackage{bm}
\usepackage{color}
\usepackage{ulem}

\definecolor{ao_en}{rgb}{0.0, 0.5, 0.0}

\defcitealias{2020MNRAS.497.3830F}{Paper I}
\defcitealias{2021MNRAS.506.5512F}{Paper II}

\title[Star cluster formation under FUV/EUV feedback]{
Far and extreme UV radiation feedback in molecular clouds and its influence on the mass and size of star clusters
}

\author[H.Fukushima \& H.Yajima]{
Hajime Fukushima$^{1}$\thanks{E-mail:fukushima@ccs.tsukuba.ac.jp},
Hidenobu Yajima$^{1}$
\\
$^{1}$Center for Computational Sciences, University of Tsukuba, Ten-nodai, 1-1-1 Tsukuba, Ibaraki 305-8577, Japan\\
}
\date{Accepted XXX. Received YYY; in original form ZZZ}

\pubyear{2020}

\begin{document}
\label{firstpage}
\pagerange{\pageref{firstpage}--\pageref{lastpage}}
\maketitle

\begin{abstract}
 We study the formation of star clusters in molecular clouds by performing three-dimensional radiation hydrodynamics simulations with far ultraviolet (FUV; $6 ~{\rm eV} \leqq h \nu \leqq 13.6 ~{\rm eV}$) and extreme ultraviolet (EUV; $h\nu \geqq 13.6~{\rm eV}$) radiative feedback. We find that the FUV feedback significantly suppresses the star formation in diffuse clouds with the initial surface densities of $\Sigma_{\rm cl} \lesssim \rm 50~M_{\odot} \; pc^{-2}$.
 In the cases of clouds with $\Sigma_{\rm cl} \sim \rm 100-200~M_{\odot} \; pc^{-2}$, the EUV feedback plays a main role and decrease the star formation efficiencies less than $0.3$.
 We show that thermal pressure from PDRs or H{\sc ii} regions disrupts the clouds and makes the size of the star clusters larger. Consequently, the clouds with the mass $M_{\rm cl} \lesssim 10^{5}~\rm M_{\odot}$ and the surface density $\Sigma_{\rm cl} \lesssim 200~\rm M_{\odot}\; pc^{-2}$ remain the star clusters with the stellar densities of $\sim 100~\rm M_{\odot}\; pc^{-3}$ that nicely match the observed open clusters in the Milky Way.
 If the molecular clouds are massive ($M_{\rm cl} \gtrsim 10^{5}~\rm M_{\odot}$) and compact ($\Sigma \gtrsim 400~\rm M_{\odot}\; pc^{-2}$), the radiative feedback is not effective and they form massive dense cluster with the stellar densities of $\sim 10^{4}~\rm M_{\odot}\; pc^{-3}$ like observed globular clusters or young massive star clusters.
 Thus, we suggest that the radiative feedback and the initial conditions of molecular clouds are key factors inducing the variety of the observed star clusters.
\end{abstract}



\section{Introduction}\label{introduction}

Revealing the formation of massive stars and their feedback mechanism is a key to understanding the cosmological star formation history.
Massive stars inject momentum and energy into the surrounding medium via radiation, stellar wind, and supernovae \citep{2019ARA&A..57..227K}.
These processes regulate star formation and induce galactic wind from galaxies \citep[e.g.,][]{2017ApJ...846...30Y, 2020arXiv201111663Y, 2019MNRAS.490.3234N}.
 In a galactic scale, the conversion from gas to stars slowly proceeds.
The depletion timescale of molecular gas in galaxies has been estimated as $\sim$ a few Gyr \citep[e.g.,][]{1998ApJ...498..541K, 2008AJ....136.2846B, 2013ApJ...772L..13M}.
Whereas, in a smaller scale,
most stars form in molecular clouds as members of star clusters initially.
The cloud lifetimes are much shorter than the depletion timescale, $\sim 10-30~{\rm Myr}$ \citep{2001ApJ...562..852H, 2009ApJS..184....1K, 2010ARA&A..48..547F}.
This indicates that the star formation in a cloud is quite inefficient and is quenched in a short timescale.
Recent studies have shown that the duration time of the star formation is less than $\sim 5~{\rm Myr}$, and the star formation efficiencies (SFEs) are less than 10 per cents \citep[e.g.,][]{2019Natur.569..519K,2020MNRAS.493.2872C}.
The timescale of the star formation is shorter than the lifetimes of OB stars.
Consequently, the pre-supernovae feedback processes, such as radiative feedback and stellar wind, are likely to play main roles in star-forming clouds \citep[e.g.,][]{2019Natur.569..519K, 2020arXiv201013788C}.

Star clusters are traditionally classified in the two categories as open star clusters and globular clusters  \citep[e.g.,][]{2019ARA&A..57..227K}.
Open clusters (OCs) are younger than $10~{\rm Gyr}$ \citep[e.g.,][]{2018A&A...614A..22P}, and their stellar mass is typically less than $10^4~M_{\odot}$ in the Milky Way.
Globular clusters (GCs) are older and more massive as $\gtrsim 10~{\rm Gyr}$ and $M_* > 10^5~M_{\odot}$ \citep[e.g.,][]{2018ARA&A..56...83B, 2018RSPSA.47470616F}.
There is an overlap between young clusters and globular clusters in mass.
Young massive star clusters (YMCs), which are massive $(> 10^4~M_{\odot})$ and high dense $(> 10^3~M_{\odot}{\rm pc^{-3}})$ \citep{2010ARA&A..48..431P}, are found in star burst galaxies \citep[e.g.,][]{2021PASJ...73S..35T, 2021PASJ...73..417T}, and in the Milky Way although they are quite rare.
Besides, there is another type of young star clusters that are massive $>10^4~M_{\odot}$ but lower stellar density $1-10~M_{\odot}{\rm pc^{-3}}$.
These are called 'leaky clusters' \citep{2009A&A...498L..37P} and are likely to form in diffuse clouds
\citep{2016ApJ...817....4F}.
Thus, recent observations have revealed that there is a large variation in observed star clusters in local galaxies.
However, the origins of these star clusters have been poorly understood.

The properties of star clusters, such as their stellar masses and densities, can change with time depending on the initial conditions after the disruption of host clouds.
Most star clusters in the Milky Way undergo the dispersal of newborn stars and remaining stars are observed as OCs \citep{2003ARA&A..41...57L}.
N-body simulations of star cluster evolution showed that more than 15-30 percent of the cloud mass should be converted into stars to keep the star cluster gravitationally bound \citep[e.g.,][]{1984ApJ...285..141L, 2001MNRAS.321..699K, 2007MNRAS.380.1589B, 2017A&A...605A.119S}.
\citet{2019MNRAS.487..364L} showed that the bound fractions of newborn stars are tightly correlated with SFEs \citep[see also][]{2021MNRAS.506.3239G, 2021MNRAS.506.5512F}.
Therefore, the SFE is a key parameter to understand the formation of star clusters. Recent simulations have indicated that the radiative feedback regulates the SFEs significantly \citep[e.g.,][]{2017MNRAS.470.3346H}.

Extreme ultraviolet (EUV; $13.6~{\rm ev} \lesssim h \nu \lesssim 1~{\rm keV}$) photons emitted from massive stars make H{\sc ii} bubbles of which the high thermal pressure can push ambient gas.
The EUV feedback finally disrupt a host cloud and quenches the star formation, resulting in the low star formation efficiencies (SFEs)
 \citep[e.g.,][]{1997ApJ...476..166W, 2002ApJ...566..302M, 2009ApJ...703.1352K, 2010ApJ...710L.142F, 2016ApJ...819..137K, 2020MNRAS.497.5061I}.
Recently, these processes have been investigated in detail by performing radiation hydrodynamics simulations (RHD)
 \citep[e.g.,][]{2010ApJ...715.1302V, 2012MNRAS.427.2852D, 2013MNRAS.430..234D, 2017MNRAS.470.3346H,2017MNRAS.471.4844G, 2017MNRAS.472.4155G, 2018ApJ...859...68K, 2019MNRAS.489.1880H, 2020MNRAS.497.4718D, 2018MNRAS.475.3511G, 2021MNRAS.506.3239G, 2020MNRAS.499..668G,2020MNRAS.495.1672B, 2021MNRAS.501.4136A, 2021PASJ..tmp...65F}.
In cases with the clouds where the EUV feedback is dominant, the SFEs increase with initial surface densities of clouds ($\Sigma$) \citep[e.g.,][]{2010ApJ...710L.142F, 2018ApJ...859...68K}.
\citet{2018ApJ...859...68K} showed that the EUV feedback limited the SFEs to less than 10 percent at $\Sigma \lesssim 10^2 ~M_{\odot}{\rm pc^{-2}}$, which is the typical surface density of GMCs in the Milky Way.
In cases of more compact clouds, the EUV feedback cannot suppress the star formation against the stronger self-gravity of the clouds \citep[e.g.,][]{2012MNRAS.427.2852D}.
\citet{2012ApJ...758L..28B} pointed out that young massive star clusters ($M_{*}>10^4~M_{\odot}$) are formed only in compact clouds whose escape velocities are larger than $10~{\rm km\;s^{-1}}$, which is the typical sound speed of the ionized gas.
Recently, we performed RHD simulations of massive star cluster formation \citep[][here after \citetalias{2021MNRAS.506.5512F}]{2021MNRAS.506.5512F}.
We showed that gravitational force from star clusters surpasses the thermal pressure of H{\sc ii} regions in compact clouds.
The SFE is enhanced up to 30 percent in such a case, forming a dense stellar core.
These clusters are categorized as YMCs.

Far ultraviolet (FUV; $6~{\rm eV} \lesssim h \nu \lesssim 13.6~{\rm eV}$)  radiation from massive stars can also be another important radiative feedback.
FUV photons can propagate beyond the ionization front and  photodissociate molecules outside the H{\sc ii} regions \citep{2005ApJ...623..917H, 2006ApJ...646..240H}.
The gas inside photodissociation regions (PDRs) is heated up to $\sim 100-10^3~{\rm K}$ via photoelectric heating  \citep[e.g.,][]{1978ApJS...36..595D, 1994ApJ...427..822B, 1999RvMP...71..173H}.
The high thermal pressure in PDRs prevents the collapse of gas, resulting in the suppression of star formation \citep[e.g.,][]{1992ApJ...385..536R, 1998ApJ...501..192D, 2015A&A...580A..49I,2020MNRAS.497.5061I}.
In addition, FUV radiation also supports to regulate the star formation in the galactic scales \citep[e.g.,][]{2020MNRAS.499.2028B}.
\citet{2015A&A...580A..49I} pointed out that the amount of photodissociated gas is much larger than ionized gas around massive stars, and the FUV feedback is the dominant mechanism to disrupt host  clouds.
On the other hand, FUV photons cannot penetrate sub-parsec filaments where are the main sites of star formation \citep[][here after \citetalias{2020MNRAS.497.3830F}]{2020MNRAS.497.3830F}.
\citet{2019ApJ...883..127N} performed the RHD simulations of gas clumps illuminated by a massive star at a distance of $0.1~{\rm pc}$.
They showed that the contribution from EUV feedback is larger than FUV feedback in evaporating gas clumps.
Besides,
clumps behind  high-density gas regions can be shielded from FUV feedback.
\citet{2019MNRAS.487.4890A} showed that
the FUV flux varies spatially because of the
shielding effects in the first $0.5~{\rm Myr}$ after massive star formation with the RHD simulations.
Thus, the inhomogenous gas structure is likely to be essential to understand the impacts of radiative feedback. However, the radiative feedback has not been understood well in the clouds with inhomogeneous clumpy gas structures.
Besides, the interplay between the FUV and EUV feedback is unclear.
Thus, in this work, we study the star cluster formation under the feedback combining FUV and EUV radiation by performing 3D RHD simulations.

In addition, in cases of low-mass star clusters, the stellar populations should be modelled carefully, because the expected number of stars cannot reproduce the modelled IMF smoothly. Therefore, following \citet{2016ApJ...819..137K}, we here take into account the stochastic sampling of the stellar population in a stellar sink particle.
We utilize the 3D RHD simulation code, \textsc{sfumato-m1},
 which is the modified version of a self-gravitational magnetohydrodynamics code with an Eulerian adaptive mesh refinement (AMR), \textsc{sfumato} \citep{2007PASJ...59..905M, 2015ApJ...801...77M}.
We adapt the radiation transfer scheme based on the moment method with M1-closure developed in \citetalias{2021MNRAS.506.5512F}.
We have newly developed the scheme of radiation sources with the stochastic stellar population.
We study the star cluster formation in the range of various cloud mass $M_{\rm cl} = 10^4-10^6~M_{\odot}$ and surface densities $\Sigma_{\rm cl} = 50-400~M_{\odot} {\rm pc^{-2}}$.

We organize the rest of the paper as the following.
In Section \ref{section_nemerical_method}, we describe the numerical method and the initial condition of the simulations.
Then, we show the results of the simulations in Section \ref{sec_results}.
In Section \ref{sec_discussion}, we discuss the implication of our results in the scenario of the star cluster formation.
Section \ref{Section_Summary_and_Discussion} are for summary and discussion.

\section{Numerical method}\label{section_nemerical_method}

\begin{table*}
    \caption{Simulation setups}
    \label{tab_init_condition}
    \centering
    \begin{tabular}{|l|c|c|c|c|c|c|c|c|c|}
      \hline \hline
      model  & $\Sigma_{\rm cl} [ M_{\odot} {\rm pc^{-1}} ]$   & $M_{\rm cl} [M_{\odot}]$ & $R_{\rm cl} [{\rm pc} ]$ & $n_{\rm ini}  [{\rm cm^{-3}} ]$ & $\alpha_{\rm 0}$  & $\sigma_0 [{\rm km /s}]$ & $v_{\rm esc} [ {\rm km/s} ]$ & $t_{\rm ff}  [{\rm Myr} ] $  & Feedback  \\
      \hline
S25M4EF & $25$ & $10^4$ & $11.3$ & $48.4$ & $1$ & $1.51$ & $2.76$ & $6.28$ & EUV+FUV \\
S25M4E & $25$ & $10^4$ & $11.3$ & $48.4$ & $1$ & $1.51$ & $2.76$ & $6.28$ & EUV \\
S25M4F & $25$ & $10^4$ & $11.3$ & $48.4$ & $1$ & $1.51$ & $2.76$ & $6.28$ & FUV \\
S25M5EF & $25$ & $10^5$ & $35.7$ & $15.3$ & $1$ & $2.69$ & $4.91$ & $11.2$ & EUV+FUV \\
S25M5E & $25$ & $10^5$ & $35.7$ & $15.3$ & $1$ & $2.69$ & $4.91$ & $11.2$ & EUV \\
S25M5F & $25$ & $10^5$ & $35.7$ & $15.3$ & $1$ & $2.69$ & $4.91$ & $11.2$ & FUV \\
S25M6EF & $25$ & $10^6$ & $113$ & $4.84$ & $1$ & $4.78$ & $8.73$ & $19.9$ & EUV+FUV \\
S25M6E & $25$ & $10^6$ & $113$ & $4.84$ & $1$ & $4.78$ & $8.73$ & $19.9$ & EUV \\
S25M6F & $25$ & $10^6$ & $113$ & $4.84$ & $1$ & $4.78$ & $8.73$ & $19.9$ & FUV \\
S50M4EF & $50$ & $10^4$ & $7.98$ & $138$ & $1$ & $1.80$ & $3.28$ & $3.73$ & EUV+FUV \\
S50M4E  & $50$ & $10^4$ & $7.98$ & $138$ & $1$ & $1.80$ & $3.28$ & $3.73$ & EUV \\
S50M4F  & $50$ & $10^4$ & $7.98$ & $138$ & $1$ & $1.80$ & $3.28$ & $3.73$ & FUV \\
S50M5EF & $50$ & $10^5$ & $25.2$ & $43.6$ & $1$  & $3.20$ & $5.84$ & $6.64$ & EUV+FUV \\
S50M5E  & $50$ & $10^5$ & $25.2$ & $43.6$ & $1$  & $3.20$ & $5.84$ & $6.64$ & EUV \\
S50M5F  & $50$ & $10^5$ & $25.2$ & $43.6$ & $1$  & $3.20$ & $5.84$ & $6.64$ & FUV \\
S50M6EF & $50$ & $10^6$ & $79.8$ & $13.8$ & $1$  & $5.69$ & $10.4$ & $11.8$ & EUV+FUV \\
S50M6E  & $50$ & $10^6$ & $79.8$ & $13.8$ & $1$  & $5.69$ & $10.4$ & $11.8$ & EUV \\
S50M6F  & $50$ & $10^6$ & $79.8$ & $13.8$ & $1$  & $5.69$ & $10.4$ & $11.8$ & FUV \\
S100M4EF & $100$ & $10^4$ & $5.64$ & $390$ & $1$  & $2.14$ & $3.90$ & $2.22$ & EUV+FUV \\
S100M4E  & $100$ & $10^4$ & $5.64$ & $390$ & $1$  & $2.14$ & $3.90$ & $2.22$ & EUV \\
S100M4F  & $100$ & $10^4$ & $5.64$ & $390$ & $1$  & $2.14$ & $3.90$ & $2.22$ & FUV \\
S100M4EFA2 & $100$ & $10^4$ & $5.64$ & $390$ & $2$  & $4.28$ & $3.90$ & $2.22$ & EUV+FUV \\
S100M4EA2 & $100$ & $10^4$ & $5.64$ & $390$ & $2$  & $4.28$ & $3.90$ & $2.22$ & EUV \\
S100M4FA2 & $100$ & $10^4$ & $5.64$ & $390$ & $2$  & $4.28$ & $3.90$ & $2.22$ & FUV \\
S100M4EFA3 & $100$ & $10^4$ & $5.64$ & $390$ & $3$  & $6.42$ & $3.90$ & $2.22$ & EUV+FUV \\
S100M4EA3 & $100$ & $10^4$ & $5.64$ & $390$ & $3$  & $6.42$ & $3.90$ & $2.22$ & EUV \\
S100M4FA3 & $100$ & $10^4$ & $5.64$ & $390$ & $3$  & $6.42$ & $3.90$ & $2.22$ & FUV \\
S100M4EFA4 & $100$ & $10^4$ & $5.64$ & $390$ & $4$  & $8.56$ & $3.90$ & $2.22$ & EUV+FUV \\
S100M4EA4 & $100$ & $10^4$ & $5.64$ & $390$ & $4$  & $8.56$ & $3.90$ & $2.22$ & EUV \\
S100M4FA4 & $100$ & $10^4$ & $5.64$ & $390$ & $4$  & $8.56$ & $3.90$ & $2.22$ & FUV \\
S100M5EF & $100$ & $10^5$ & $17.8$ & $123$ & $1$  & $3.80$ & $6.94$ & $3.95$ & EUV+FUV \\
S100M5E  & $100$ & $10^5$ & $17.8$ & $123$ & $1$  & $3.80$ & $6.94$ & $3.95$ & EUV \\
S100M5F  & $100$ & $10^5$ & $17.8$ & $123$ & $1$  & $3.80$ & $6.94$ & $3.95$ & FUV \\
S100M6EF & $100$ & $10^6$ & $56.4$ & $39.0$ & $1$  & $6.76$ & $12.3$ & $7.02$ & EUV+FUV \\
S100M6E  & $100$ & $10^6$ & $56.4$ & $39.0$ & $1$  & $6.76$ & $12.3$ & $7.02$ & EUV \\
S100M6F  & $100$ & $10^6$ & $56.4$ & $39.0$ & $1$  & $6.76$ & $12.3$ & $7.02$ & FUV \\
S200M4EF & $200$ & $10^4$ & $3.99$ & $1100$ & $1$  & $2.54$ & $4.64$ & $1.32$ & EUV+FUV \\
S200M4E  & $200$ & $10^4$ & $3.99$ & $1100$ & $1$  & $2.54$ & $4.64$ & $1.32$ & EUV \\
S200M4F  & $200$ & $10^4$ & $3.99$ & $1100$ & $1$  & $2.54$ & $4.64$ & $1.32$ & FUV \\
S200M5EF & $200$ & $10^5$ & $12.6$ & $349$ & $1$  & $4.52$ & $8.26$ & $2.35$ & EUV+FUV \\
S200M5E  & $200$ & $10^5$ & $12.6$ & $349$ & $1$  & $4.52$ & $8.26$ & $2.35$ & EUV \\
S200M5F  & $200$ & $10^5$ & $12.6$ & $349$ & $1$  & $4.52$ & $8.26$ & $2.35$ & FUV \\
S200M6EF & $200$ & $10^6$ & $39.9$ & $110$ & $1$  & $8.04$ & $14.7$ & $4.18$ & EUV+FUV \\
S200M6E  & $200$ & $10^6$ & $39.9$ & $110$ & $1$  & $8.04$ & $14.7$ & $4.18$ & EUV \\
S200M6F  & $200$ & $10^6$ & $39.9$ & $110$ & $1$  & $8.04$ & $14.7$ & $4.18$ & FUV \\
S400M4EF & $400$ & $10^4$ & $2.82$ & $3120$ & $1$  & $3.02$ & $5.52$ & $0.790$ & EUV+FUV \\
S400M4E  & $400$ & $10^4$ & $2.82$ & $3120$ & $1$  & $3.02$ & $5.52$ & $0.790$ & EUV \\
S400M4F  & $400$ & $10^4$ & $2.82$ & $3120$ & $1$  & $3.02$ & $5.52$ & $0.790$ & FUV \\
S400M5EF & $400$ & $10^5$ & $8.92$ & $986$ & $1$  & $5.38$ & $9.82$ & $1.40$ & EUV+FUV \\
S400M5E  & $400$ & $10^5$ & $8.92$ & $986$ & $1$  & $5.38$ & $9.82$ & $1.40$ & EUV \\
S400M5F  & $400$ & $10^5$ & $8.92$ & $986$ & $1$  & $5.38$ & $9.82$ & $1.40$ & FUV \\
S400M6EF & $400$ & $10^6$ & $28.2$ & $312$ & $1$  & $9.57$ & $17.5$ & $2.48$ & EUV+FUV \\
S400M6E  & $400$ & $10^6$ & $28.2$ & $312$ & $1$  & $9.57$ & $17.5$ & $2.48$ & EUV \\
S400M6F  & $400$ & $10^6$ & $28.2$ & $312$ & $1$  & $9.57$ & $17.5$ & $2.48$ & FUV \\

      \hline
    \end{tabular}
    \begin{minipage}{1\hsize}
     Notes. Column 1: model names, Column 2: surfaces densities, Column 3: cloud masses, Column 4: cloud radii, Column 5: initial number densities, Column 6: virial parameters, Column 7:three-dimensional velocity dispersion, Column 8: escape velocities, Column 9: free fall times, Column 10: Feedback mechanisms
    \end{minipage}
  \end{table*}

We perform RHD simulations with SFUMATO-M1 \citepalias{2021MNRAS.506.5512F}, the modified version of self-gravitational magnetohydrodynamics code with AMR, SFUMATO \citep{2007PASJ...59..905M}.
We solve hydrodynamics equations coupling with radiation transfer based on the moment method with M1 closure and non-equilibrium chemistry.
The chemistry solver was developed in \citet{2020ApJ...892L..14S}.
Here, we add the chemical network of  CO formation \citep{1997ApJ...482..796N} and the oxygen ions in H{\sc ii} regions \citep[OII, OIII, also see ][]{2020MNRAS.497..829F}.
In the simulations, we put sink particles where the local density is higher than a threshold value $\rho_{\rm thr}$.
The model of the sink particle is developed in \citet{2015ApJ...801...77M}.
The threshold density is set as $\rho_{\rm thr} = 9.3 \times 10^{-19} (R_{\rm cl}/20~{\rm pc})^{-2} ~{\rm g \, \rm cm^{-3}}$ \citep{2013ApJS..204....8G}.
We consider the sink radius as $r_{\rm sink} = 2 \Delta x$ where $\Delta x$ is the cell size at the maximum refinement level.
While the model of the sink particle is almost similar to our previous work \citetalias{2021MNRAS.506.5512F}, in this work, we newly take into account the stochastic formation of multiple stars in a sink particle. This induces variations of feedback strength from sink particles even if their particle mass is the same.

\subsection{Stochastic stellar population}\label{section_stochastic_stellar_population}

The emissivity of a star cluster depends on the population of member stars.
When the number of stars is not enough to follow the modelled initial mass function (IMF), the mass-to-luminosity ratio of the star cluster fluctuates according to the stellar mass distribution.
\citet{2016ApJ...819..137K} introduced the stochastic stellar population model with the various stellar cluster masses, using the SLUG code \citep{2012ApJ...745..145D, 2015MNRAS.452.1447K}.
They showed that the photon production rate per stellar mass is comparable to the IMF-averaged value only if the total stellar mass exceeds $10^4~M_{\odot}$.
Therefore, in cases of low-mass star clusters, the emissivity can change stochastically
and the SFEs are likely to be different even for the same initial condition.
\citet{2019MNRAS.488.2970G} showed that SFEs vary by a factor of $2-3$ with the model of radiation sources in star cluster formation.

Here, in the development of the sub-grid model of the stochastic stellar populations,
we adopt a method similar to SLUG code developed by \citet{2012ApJ...745..145D} and \citet{2015MNRAS.452.1447K}.
We divide the stellar masses into 100 bins between $0.1$ and $120~M_{\odot}$ in a log scale.
We stochastically distribute stars into these bins based on the probability weighted with Chabier IMF \citep{2003PASP..115..763C} when a new sink particle is formed.
Then, the sink particle mass can grow with the gas accretion.
The sink masses are typically distributed in the range from $\sim 1 ~M_{\odot}$ to $\sim 10^3~M_{\odot}$ in the simulations.
The birth of the star is delayed if the available gas mass in a sink particle is lower than the stellar mass chosen stochastically.
We record the initial stellar mass and the formation time of each star when they are distributed.

There is a time interval between the formation of massive stars and when they start to emit UV photons efficiently. Therefore,
we consider the stellar evolution to estimate the radiative properties for each stellar bin.
We adapt the PARSEC tracks \citep{2012MNRAS.427..127B, 2014MNRAS.445.4287T, 2014MNRAS.444.2525C, 2015MNRAS.452.1068C, 2017ApJ...835...77M, 2019MNRAS.485.5666P, 2020MNRAS.498.3283P} to obtain the evolution of luminosity and effective temperatures ($L_{*}$ and $T_{\rm eff}$).
Because the stellar atmosphere absorbs UV photons, UV photon emissivity deviates from the estimate with the black-body spectrum \citep[e.g., ][]{1979ApJS...40....1K, 2020ApJS..250...13K}.
In this study, We use the SED models in \citet{1997A&AS..125..229L} and \citet{2019A&A...621A..85H} for OB-stars.
With these SED models, we pre-calculate the radiative properties, such as the EUV and FUV photon emissivities.

\begin{figure}
    \begin{center}
    	\includegraphics[width=\columnwidth]{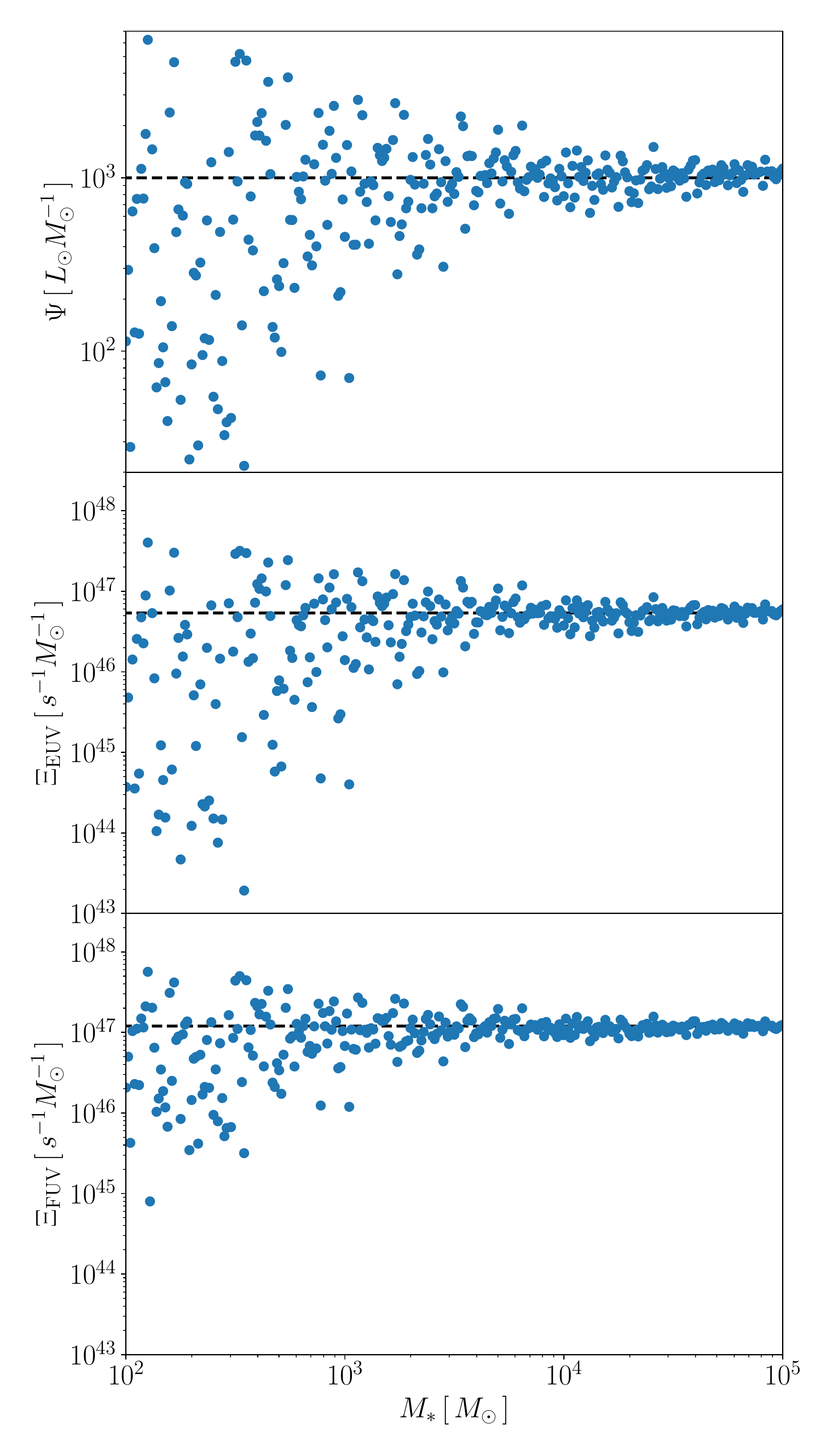}
    \end{center}
    \caption{
    The radiative properties of star clusters in the stochastic stellar population model.
    Each panel represents the light-to-mass ratio ($\Phi$), the number of EUV and FUV photons per stellar mass unit ($\Xi_{\rm EUV}$ and $\Xi_{\rm FUV}$) from top to bottom.
    The ages of stars are set at $10^{6}~{\rm yr}$.
    In each panel, dashed lines show the IMF-averaged values.
}
    \label{fig_mass_rad}
\end{figure}

Figure \ref{fig_mass_rad} shows the mass-to-luminosity rations ($\Psi$), the emission rates of EUV and FUV photons per stellar mass unit ($\Xi_{\rm EUV}$ and $\Xi_{\rm FUV}$) when the stellar ages are $10^{6}~{\rm yr}$.
The dashed lines represent the IMF-averaged values.
In low-mass stellar clusters with $M_* < 10^{4}~M_{\odot}$,
there is a large dispersion due to the stochastic sampling.
Note that the values tend to be smaller than the IMF-averaged values.
This is because the expected number of massive stars is below unity in the low-mass range $M_* < 1.2 \times 10^{2}~M_{\odot}$.
As the cluster mass increases, the stellar population can be fully sampled.  At $M_* > 10^{4}~M_{\odot}$, the emissivities are almost the same as the IMF-averaged ones (see more detail in Section 3.2).

\subsection{Radiative transfer and thermal processes}\label{Section_RT_Thermal}

We take into account the thermal processes of chemical reactions, $\rm H_2$ and metal line cooling, and energy transfer between gas and dust grains
(see \citetalias{2021MNRAS.506.5512F}).
In addition to our previous work, we here consider (1) the photoelectric heating, (2) heating from $\rm H_{2}$ UV pumping, and (3) ionization of cosmic rays as described below.

We calculate radiative transfer with M1-closure method \citep[e.g.,][]{2013MNRAS.436.2188R, 2015MNRAS.449.4380R,2019MNRAS.485..117K} that was developed in \citetalias{2021MNRAS.506.5512F}.
We set  four frequency bins for (1) extreme ultraviolet (EUV;$13.6~{\rm eV} < h \nu$), (2) Lyman-Werner (LW; $11.2 ~{\rm eV} < h \nu < 13.6~{\rm eV}$), (3)  far-ultraviolet (FUV; $6~{\rm eV} < h \nu < 13.6~{\rm eV}$), and (4) infrared (IR) photons.
The EUV, LW, and FUV photons are directly emitted from stars.
We consider the dust absorption to these photons and also the absorption by neutral hydrogen to EUV photons. The photon absorption is also considered in the hydrodynamics momentum equation as the radiation pressure force.
The UV photons can change the chemical abundances.
The EUV photons cause the photoionization of neutral hydrogen,
while the LW photons induce photodissociation of ${\rm H_2}$ and CO molecules.
We include the self-shielding of molecules at high-density regions where the column density of $\rm H_2$ is higher than $10^{14}~{\rm cm^{-2}}$  \citep{1999RvMP...71..173H}.
We adopt the fitting function of the self-shielding for hydrogen molecules derived in  \citet{2019MNRAS.484.2467W}, which is the modified one of \citet{1996ApJ...468..269D}.
For the photodissociation and the shelf-shielding rates of CO molecules,
we use the fitting functions derived in \citet{1996A&A...311..690L}.
The radiative transfer calculations with the moment-based method cannot follow the trajectories of photons.
Thus, we need to estimate the column density of molecules only from the local physical variables.
\citet{2011MNRAS.418..838W} indicated that the local Sobolev length defined as
\begin{align}
    L_{\rm sol} = \frac{v_{\rm th}}{|dv/ds | } \label{eq_solbolev},
\end{align}
is suitable to estimate the local column densities.
In Equation \eqref{eq_solbolev}, $v_{\rm th}$ is the thermal velocity of each molecule, and $|dv/ds|$ is the velocity gradient.
In a static media, the Sobolev length becomes infinite because of non-velocity gradient.
We set the column density estimated from a local Jeans length $\lambda_{\rm J}$ as the maximum values to avoid overestimating the self-shielding factor.
Besides, we include the UV pumping and photodisocciation of $\rm H_2$ as the FUV heating processes with the formulae derived in \citet{1979ApJS...41..555H}.

Using the local gas temperature $T$, the electron number density $n_{\rm e}$, and the strength of FUV radiation field $G_{\rm fuv}$, we estimate the rates of
photoelectric heating and recombination cooling with the functions derived in \citet{1994ApJ...427..822B}. Here, $G_{\rm fuv}$ is described in the unit of Habing flux $(1.6 \times 10^{-3}~{\rm erg \, {\rm cm^{-2}} \, {\rm s^{-1}}})$.

We adopt a method similar to \citet{2017ApJ...843...38G} for cosmic-ray ionization.
The primary and secondary ionization processes are taken into account.
Here, we use a simple model of \citet{1974ApJ...193...73G} where the total ionization rate for atomic hydrogen is given as 1.5 times of the primary ionization.
Besides, we assume that the total ionization rate per hydrogen atom is the same for molecular hydrogen for simplicity.
The primary cosmic-ray ionization rate is given as $\xi_{\rm H} = 2 \times 10^{-16} ~{\rm s^{-1}} {\rm H}^{-1}$ \citep{2007ApJ...671.1736I}.
The heating rates of cosmic-ray ionization are estimated by \citet{2011piim.book.....D} and \citet{2014MNRAS.437.1662K} for atomic hydrogen and hydrogen molecule.

\subsection{Initial conditions}\label{Section_initial_condition}

\begin{table*}
    \caption{Simulation results}
    \label{tab_results}
    \centering
    \begin{tabular}{|l|c|c|c|c|c|c|c|c|}
      \hline \hline
       model & $\epsilon_*$ & $t_{\rm life}^{\rm a} \, [ {\rm Myr} ]$ & $t_{\rm dr}^{\rm b} \, [ {\rm Myr} ]$ & $\bar{\Psi} / \Psi_0 ^{\rm c}$ & $\bar{\Xi}_{\rm EUV} / \Xi_{\rm EUV, 0}^{\rm d}$ & $\bar{\Xi}_{\rm FUV} / \Xi_{\rm FUV, 0}^{\rm e}$ & $f_{\rm bd}$ & $r_{\rm h} \, [{\rm pc}]$ \\ \hline
S25M4EF & $0.1$ & $9.08 \, (1.45~t_{\rm ff} )$ & $3.38 \, (0.54~t_{\rm ff} )$ & $0.12$ & $0.05$ & $0.15$ & $0.42$ & $1.01$ \\
S25M4E & $0.08$ & $7.57 \, (1.21~t_{\rm ff} )$ & $1.95 \, (0.31~t_{\rm ff} )$ & $0.2$ & $0.1$ & $0.25$ & $0.32$ & $0.67$ \\
S25M4F & $0.14$ & $9.49 \, (1.51~t_{\rm ff} )$ & $3.67 \, (0.58~t_{\rm ff} )$ & $0.19$ & $0.11$ & $0.23$ & $0.63$ & $0.8$ \\
S25M5EF & $0.06$ & $17.38 \, (1.56~t_{\rm ff} )$ & $8.26 \, (0.74~t_{\rm ff} )$ & $0.36$ & $0.19$ & $0.44$ & $0.18$ & $2.62$ \\
S25M5E & $0.07$ & $18.78 \, (1.68~t_{\rm ff} )$ & $9.64 \, (0.86~t_{\rm ff} )$ & $0.41$ & $0.22$ & $0.49$ & $0.03$ & $0.07$ \\
S25M5F & $0.09$ & $16.49 \, (1.48~t_{\rm ff} )$ & $7.07 \, (0.63~t_{\rm ff} )$ & $0.53$ & $0.34$ & $0.61$ & $0.62$ & $2.01$ \\
S25M6EF & $0.04$ & $28.05 \, (1.41~t_{\rm ff} )$ & $13.47 \, (0.68~t_{\rm ff} )$ & $0.72$ & $0.43$ & $0.82$ & $0.04$ & $0.9$ \\
S25M6E & $0.05$ & $32.09 \, (1.62~t_{\rm ff} )$ & $17.52 \, (0.88~t_{\rm ff} )$ & $0.64$ & $0.38$ & $0.71$ & $0.0$ & $0.0$ \\
S25M6F & $0.09$ & $29.89 \, (1.51~t_{\rm ff} )$ & $13.91 \, (0.7~t_{\rm ff} )$ & $0.74$ & $0.48$ & $0.8$ & $0.2$ & $4.67$ \\
S50M4EF & $0.18$ & $6.58 \, (1.76~t_{\rm ff} )$ & $3.18 \, (0.85~t_{\rm ff} )$ & $0.18$ & $0.08$ & $0.23$ & $0.81$ & $1.54$ \\
S50M4E & $0.22$ & $6.37 \, (1.71~t_{\rm ff} )$ & $2.95 \, (0.79~t_{\rm ff} )$ & $0.19$ & $0.11$ & $0.22$ & $0.73$ & $1.76$ \\
S50M4F & $0.27$ & $6.61 \, (1.77~t_{\rm ff} )$ & $3.13 \, (0.84~t_{\rm ff} )$ & $0.36$ & $0.25$ & $0.4$ & $0.93$ & $0.29$ \\
S50M5EF & $0.06$ & $11.95 \, (1.8~t_{\rm ff} )$ & $6.81 \, (1.03~t_{\rm ff} )$ & $0.74$ & $0.5$ & $0.83$ & $0.04$ & $0.76$ \\
S50M5E & $0.09$ & $11.04 \, (1.66~t_{\rm ff} )$ & $5.8 \, (0.87~t_{\rm ff} )$ & $0.7$ & $0.46$ & $0.79$ & $0.08$ & $1.15$ \\
S50M5F & $0.23$ & $11.13 \, (1.68~t_{\rm ff} )$ & $5.43 \, (0.82~t_{\rm ff} )$ & $0.79$ & $0.59$ & $0.82$ & $0.84$ & $1.61$ \\
S50M6EF & $0.07$ & $18.03 \, (1.53~t_{\rm ff} )$ & $9.32 \, (0.79~t_{\rm ff} )$ & $0.8$ & $0.54$ & $0.87$ & $0.0$ & $0.0$ \\
S50M6E & $0.08$ & $17.48 \, (1.48~t_{\rm ff} )$ & $8.75 \, (0.74~t_{\rm ff} )$ & $0.74$ & $0.48$ & $0.83$ & $0.02$ & $1.89$ \\
S50M6F & $0.37$ & $18.98 \, (1.61~t_{\rm ff} )$ & $8.21 \, (0.7~t_{\rm ff} )$ & $0.87$ & $0.45$ & $0.66$ & $0.98$ & $2.66$ \\
S100M4EF & $0.19$ & $3.34 \, (1.51~t_{\rm ff} )$ & $1.38 \, (0.62~t_{\rm ff} )$ & $0.26$ & $0.16$ & $0.3$ & $0.57$ & $0.49$ \\
S100M4E & $0.26$ & $3.76 \, (1.69~t_{\rm ff} )$ & $1.77 \, (0.8~t_{\rm ff} )$ & $0.26$ & $0.15$ & $0.31$ & $0.79$ & $1.0$ \\
S100M4F & $0.48$ & $3.79 \, (1.71~t_{\rm ff} )$ & $1.73 \, (0.78~t_{\rm ff} )$ & $0.29$ & $0.21$ & $0.31$ & $0.99$ & $0.14$ \\
S100M4EFA2 & $0.14$ & $4.76 \, (2.14~t_{\rm ff} )$ & $2.81 \, (1.27~t_{\rm ff} )$ & $0.9$ & $0.87$ & $0.83$ & $0.67$ & $2.46$ \\
S100M4EA2 & $0.18$ & $4.42 \, (1.99~t_{\rm ff} )$ & $2.44 \, (1.1~t_{\rm ff} )$ & $0.72$ & $0.75$ & $0.61$ & $0.74$ & $2.49$ \\
S100M4FA2 & $0.28$ & $5.24 \, (2.36~t_{\rm ff} )$ & $3.18 \, (1.43~t_{\rm ff} )$ & $0.31$ & $0.14$ & $0.4$ & $0.96$ & $0.42$ \\
S100M4EFA3 & $0.12$ & $5.14 \, (2.31~t_{\rm ff} )$ & $3.13 \, (1.41~t_{\rm ff} )$ & $0.05$ & $0.01$ & $0.07$ & $0.58$ & $0.81$ \\
S100M4EA3 & $0.13$ & $5.26 \, (2.37~t_{\rm ff} )$ & $3.24 \, (1.46~t_{\rm ff} )$ & $0.85$ & $0.73$ & $0.85$ & $0.61$ & $2.93$ \\
S100M4FA3 & $0.17$ & $6.1 \, (2.75~t_{\rm ff} )$ & $3.99 \, (1.8~t_{\rm ff} )$ & $0.2$ & $0.13$ & $0.22$ & $0.9$ & $1.64$ \\
S100M4EFA4 & $0.06$ & $6.51 \, (2.93~t_{\rm ff} )$ & $4.45 \, (2.0~t_{\rm ff} )$ & $0.06$ & $0.01$ & $0.08$ & $0.51$ & $1.67$ \\
S100M4EA4 & $0.08$ & $7.1 \, (3.2~t_{\rm ff} )$ & $4.98 \, (2.24~t_{\rm ff} )$ & $0.13$ & $0.05$ & $0.17$ & $0.46$ & $1.76$ \\
S100M4FA4 & $0.07$ & $6.33 \, (2.85~t_{\rm ff} )$ & $4.26 \, (1.92~t_{\rm ff} )$ & $0.08$ & $0.01$ & $0.11$ & $0.52$ & $2.08$ \\
S100M5EF & $0.16$ & $6.81 \, (1.73~t_{\rm ff} )$ & $3.65 \, (0.92~t_{\rm ff} )$ & $0.67$ & $0.5$ & $0.72$ & $0.38$ & $4.19$ \\
S100M5E & $0.17$ & $6.82 \, (1.73~t_{\rm ff} )$ & $3.68 \, (0.93~t_{\rm ff} )$ & $0.86$ & $0.7$ & $0.88$ & $0.19$ & $2.53$ \\
S100M5F & $0.53$ & $6.89 \, (1.75~t_{\rm ff} )$ & $3.37 \, (0.85~t_{\rm ff} )$ & $0.88$ & $0.67$ & $0.74$ & $0.99$ & $0.6$ \\
S100M6EF & $0.13$ & $12.44 \, (1.77~t_{\rm ff} )$ & $7.27 \, (1.03~t_{\rm ff} )$ & $0.81$ & $0.59$ & $0.87$ & $0.04$ & $8.24$ \\
S100M6E & $0.14$ & $12.4 \, (1.77~t_{\rm ff} )$ & $7.24 \, (1.03~t_{\rm ff} )$ & $0.8$ & $0.58$ & $0.86$ & $0.01$ & $2.73$ \\
S100M6F & $0.48$ & $11.48 \, (1.63~t_{\rm ff} )$ & $5.29 \, (0.75~t_{\rm ff} )$ & $0.91$ & $0.75$ & $0.92$ & $0.97$ & $2.04$ \\
S200M4EF & $0.35$ & $2.33 \, (1.77~t_{\rm ff} )$ & $1.17 \, (0.89~t_{\rm ff} )$ & $0.31$ & $0.21$ & $0.32$ & $0.91$ & $0.61$ \\
S200M4E & $0.42$ & $2.6 \, (1.97~t_{\rm ff} )$ & $1.43 \, (1.08~t_{\rm ff} )$ & $0.62$ & $0.53$ & $0.53$ & $0.96$ & $0.28$ \\
S200M4F & $0.59$ & $2.75 \, (2.08~t_{\rm ff} )$ & $1.54 \, (1.17~t_{\rm ff} )$ & $0.61$ & $0.54$ & $0.56$ & $0.97$ & $0.16$ \\
S200M5EF & $0.24$ & $4.78 \, (2.04~t_{\rm ff} )$ & $2.93 \, (1.25~t_{\rm ff} )$ & $0.89$ & $0.77$ & $0.86$ & $0.82$ & $2.43$ \\
S200M5E & $0.28$ & $4.62 \, (1.97~t_{\rm ff} )$ & $2.75 \, (1.17~t_{\rm ff} )$ & $0.83$ & $0.71$ & $0.81$ & $0.82$ & $2.93$ \\
S200M5F & $0.66$ & $4.5 \, (1.92~t_{\rm ff} )$ & $2.44 \, (1.04~t_{\rm ff} )$ & $0.7$ & $0.48$ & $0.57$ & $1.0$ & $0.49$ \\
S200M6EF & $0.25$ & $8.32 \, (1.99~t_{\rm ff} )$ & $5.04 \, (1.21~t_{\rm ff} )$ & $0.91$ & $0.75$ & $0.92$ & $0.86$ & $7.55$ \\
S200M6E & $0.32$ & $10.26 \, (2.46~t_{\rm ff} )$ & $6.98 \, (1.67~t_{\rm ff} )$ & $0.77$ & $0.55$ & $0.82$ & $0.92$ & $5.22$ \\
S200M6F & $0.6$ & $7.19 \, (1.72~t_{\rm ff} )$ & $3.56 \, (0.85~t_{\rm ff} )$ & $0.92$ & $0.82$ & $0.9$ & $0.99$ & $1.5$ \\
S400M4EF & $0.49$ & $1.58 \, (2.01~t_{\rm ff} )$ & $0.89 \, (1.14~t_{\rm ff} )$ & $1.13$ & $1.25$ & $0.86$ & $0.97$ & $0.27$ \\
S400M4E & $0.47$ & $1.55 \, (1.98~t_{\rm ff} )$ & $0.87 \, (1.11~t_{\rm ff} )$ & $0.66$ & $0.59$ & $0.6$ & $0.97$ & $0.27$ \\
S400M4F & $0.66$ & $1.93 \, (2.46~t_{\rm ff} )$ & $1.23 \, (1.57~t_{\rm ff} )$ & $0.97$ & $0.98$ & $0.78$ & $1.0$ & $0.17$ \\
S400M5EF & $0.37$ & $2.7 \, (1.94~t_{\rm ff} )$ & $1.58 \, (1.13~t_{\rm ff} )$ & $0.81$ & $0.73$ & $0.7$ & $0.92$ & $1.16$ \\
S400M5E & $0.38$ & $2.93 \, (2.1~t_{\rm ff} )$ & $1.81 \, (1.3~t_{\rm ff} )$ & $0.61$ & $0.53$ & $0.6$ & $0.89$ & $1.03$ \\
S400M5F & $0.67$ & $2.75 \, (1.97~t_{\rm ff} )$ & $1.56 \, (1.11~t_{\rm ff} )$ & $0.88$ & $0.69$ & $0.64$ & $1.0$ & $0.41$ \\
S400M6EF & $0.59$ & $8.37 \, (3.37~t_{\rm ff} )$ & $6.27 \, (2.52~t_{\rm ff} )$ & $0.93$ & $0.7$ & $0.96$ & $0.98$ & $2.37$ \\
S400M6E & $0.62$ & $8.31 \, (3.35~t_{\rm ff} )$ & $6.25 \, (2.52~t_{\rm ff} )$ & $0.88$ & $0.65$ & $0.94$ & $0.96$ & $2.16$ \\
S400M6F & $0.7$ & $4.7 \, (1.89~t_{\rm ff} )$ & $2.65 \, (1.07~t_{\rm ff} )$ & $0.92$ & $0.86$ & $0.86$ & $0.99$ & $1.46$ \\
      \hline
    \end{tabular}
    \begin{minipage}{1\hsize}
     Notes. Column 1: model names, Column 2: star formation efficiencies, Column 3: lifetime of clouds, Column 4: duration time of star formation, Column 5: light-to-mass ratio $(\bar{\Psi})$, Column 6: emission rate of EUV photons per unit stellar mass, Column 7: emission rate of FUV photons per unit stellar mass, Column 8: bound fractions of stars, Column 9: half mass radii at the end of star formation.

     $^{\rm a}$ The cloud lifetime is defined as the period from the staring time of the simulations to the time when the total stellar mass reaches 95 per cent of the final one.

     $^{\rm b}$ The duration time of the star formation is defined as the time required for the stellar mass to increases from 5 per cent to 95 percent of the final stellar mass.

     $^{\rm c,d,e}$ $\Phi_0$, $\Xi_{\rm EUV, 0}$ and $\Xi_{\rm FUV, 0}$ denote the IMF-averaged values.
    \end{minipage}
  \end{table*}

We study the evolutions of the clouds with the masses $M_{\rm cl} = 10^4$, $10^5$, and $10^6~M_{\odot}$ and the surface densities ranging $\Sigma_{\rm cl} = 50-400~M_{\odot}{\rm pc^{-2}}$ as summarized in Table \ref{tab_init_condition}.
The maximum refinement level is fixed at $l_{\rm max} = 4$.
The minimum cell size is $\Delta x = 0.059 ~{\rm pc} (R_{\rm cl}/20~{\rm pc})$ where $R_{\rm cl}$ is the cloud radius.
For instance, the minimum cell sizes are $0.017$ and $0.17~{\rm pc}$ in the cases with $(M_{\rm cl}, \Sigma_{\rm cl}) = (10^4~M_{\odot}, 10^2~M_{\odot}{\rm pc^{-2}})$ and $(10^6~M_{\odot}, 10^2~M_{\odot}{\rm pc^{-2}})$.
As shown in \citetalias{2020MNRAS.497.3830F} and  \citetalias{2021MNRAS.506.5512F}, the SFE mainly depends on the surface densities $\Sigma_{\rm cl}$ when the star formation is regulated by the radiative feedback \citep[see also, ][]{2010ApJ...710L.142F, 2016ApJ...829..130R, 2017MNRAS.471.4844G, 2018MNRAS.475.3511G, 2018ApJ...859...68K, 2019MNRAS.489.1880H}.
In most cases, we showed that photoionization feedback is the dominant mechanism to disrupt clouds.
As shown in \citetalias{2021MNRAS.506.5512F}, if the clouds are massive and compact as $(M_{\rm cl}, \Sigma_{\rm cl})=(10^6~M_{\odot}, 400~M_{\odot}{\rm pc^{-2}})$, the photoionization feedback cannot overcome the deep gravitational potential well, resulting in the formation of high-density stellar cores.
Consequently, the SFEs exceed 0.3, and gravitationally bound star clusters form.

In this study, we focus on the heating processes induced by EUV and FUV photons, referred to as the EUV and FUV feedback.
In the fiducial models of each cloud, we include both EUV and FUV feedback.
We label these models as S100M4EF where "E" and "F" mean the EUV and FUV feedback.
We perform the additional simulations only with EUV or FUV feedback to investigate their effect on the star cluster formation.
The processes related with EUV or FUV feedback are listed below:
\begin{itemize}
    \item {\it EUV feedback:} photoionization of neutral hydrogen.
    \item {\it FUV feedback:} photoelectric heating, dissociation heating of $\rm H_2$, and UV pumping of $\rm H_2$.
\end{itemize}
Even if either feedback is turned off, the radiation pressure from both wavelength ranges keeps being considered.

As in \citetalias{2020MNRAS.497.3830F} and \citetalias{2021MNRAS.506.5512F}, we take turbulent motions into account in the initial conditions.
We assume that the velocity power spectrum is $P(k) \propto k^{-4}$ where $k$ is the wavenumber.
The amplitude of the turbulent motion is characterized with the virial parameter defined as
\begin{align}
    \alpha_{0} = \frac{2E_{\rm kin}}{ | E_{\rm grav}|} = \frac{5 \sigma_0^2 R_{\rm cl}}{3 G M_{\rm cl}}, \label{eq_alpha_vir}
\end{align}
where $\sigma_0$, $E_{\rm kin}$, and $E_{\rm grav}$ are the 3D velocity dispersion, kinetic and graviational energy.
In fiducial cases, we adapt $\alpha_0 = 1$.
The velocity field is generated with a random number.
The SFEs vary with the different choice of the seed of the random number \citep[e.g.,][]{2021MNRAS.506.3239G, 1994ApJ...427..822B, 2021ApJ...911..128K}.
Here, we focus on investigating the effects of each radiative feedback.
We rescale the same velocity fields.
Besides, the virial parameter varies on the formation site of the cloud and the observational scales \citep[e.g.,][]{2010ApJ...723..492R, 2016ApJ...831...16L, 2018ApJ...860..172S, 2021arXiv210705750E}, and the SFEs decreases with the higher virial parameter \citep[e.g.,][]{2021ApJ...911..128K, 2021MNRAS.506.5512F}.
To investigate the effects of the EUV and FUV feedback in larger virial parameters, we additionally perform the cases with $\alpha_0 = 2, 3$, and $4$ for the clouds of $(\Sigma_{\rm cl}, M_{\rm cl}) = (10^2~M_{\odot}{\rm pc^{-2}}, 10^4~M_{\odot})$.

\section{Results} \label{sec_results}

We first study the effects of EUV/FUV feedback in the star cluster formation in  Section \ref{Sec_feducial_model}.
In Section \ref{Sect_Dep_virial}, we discuss the dependence of SFEs on the virial parameters.
In Section \ref{Sec_stochastic_stellar_pop}, we investigate the impacts of the stochastic stellar population model.
In Table \ref{tab_init_condition}, we summarize the results of our simulations.

\subsection{Star cluster formation and cloud dispersal under EUV and FUV feedback}\label{Sec_feducial_model}

\subsubsection{Star cluster formation in a fiducial model}\label{Sec_evolv_fiducial_model}

\begin{figure*}
    \begin{center}
    	\includegraphics[width=170mm]{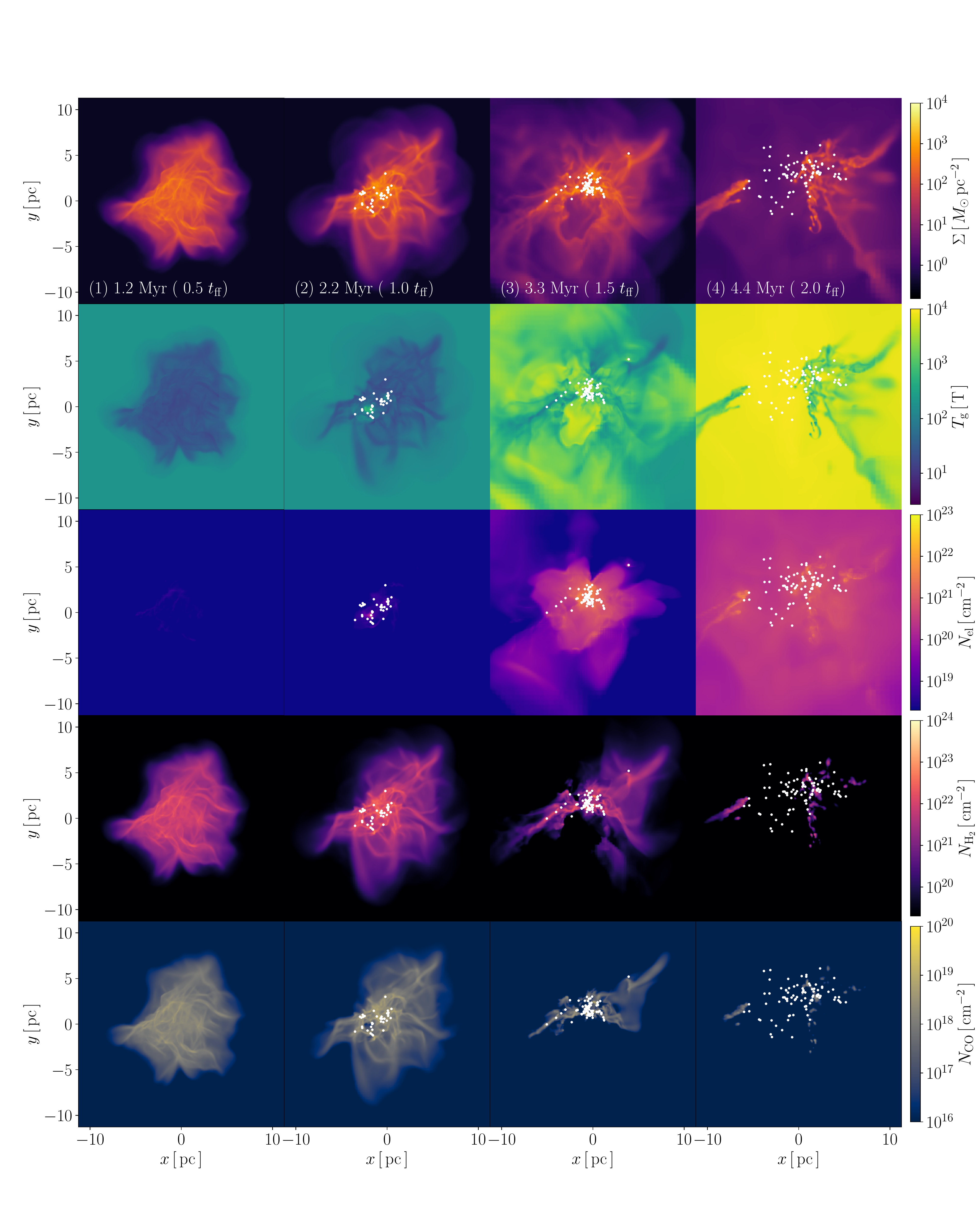}
    \end{center}
    \caption{
    Star cluster formation and cloud dispersal in the case with $M_{\rm cl} = 10^4~M_{\odot}$ and $\Sigma_{\rm cl}=100~\rm M_{\odot} \; pc^{-2}$.
    Each panel shows the surface density ($\Sigma$), density weighted gas temperature ($T_{\rm g}$), the column density of electron ($N_{\rm el}$), hydrogen molecule ($N_{\rm H_2}$), and CO molecules ($N_{\rm CO}$) from top to bottom.
    The white dots represent the positions of star cluster particles.
}
    \label{fig_sigma100m4}
\end{figure*}

We present the model of S100M4EF as the fiducial model.
Figure \ref{fig_sigma100m4} shows the time evolution of the cloud.
The entire evolution is almost the same as in \citetalias{2020MNRAS.497.3830F} and \citetalias{2021MNRAS.506.5512F}.
The turbulent motions compress the gas and induce the filamentary structures where stars newly form (see Fig. \ref{fig_sigma100m4}-1 and 2).
Once massive stars born, H{\sc ii} regions start to expand outward (see Fig. \ref{fig_sigma100m4}-3).
Consequently, the filaments are gradually evaporated due to the photoionization feedback.
Finally, most parts of the cloud dispersed at $t\sim 2~t_{\rm ff}$ as shown in Figure \ref{fig_sigma100m4}-(4), while
the pillar structures of neutral gas remain around the star cluster.

EUV radiation from massive stars heats up and photoionizes the cloud.
At $t=1.5~{\rm t_{\rm ff}}$, the ionization fronts reach $\sim 5 ~{\rm pc}$ from the star cluster.
The temperature of the ionized gas becomes $\sim 10^4~{\rm K}$, and the high thermal pressure pushes the ambient gas.
Besides, FUV photons also contribute to heating the gas.
FUV photons can propagate beyond the ionization fronts and heat the gas to $\gtrsim 10^3~{\rm K}$ in the low-density regions.
On the other hand,  high-density filaments are shielded from both EUV and FUV feedback by dust attenuation, and they remain at temperature $\lesssim 10^2 ~{\rm K}$ as shown in Figure \ref{fig_sigma100m4}-(3) \citep[see also,][]{2020MNRAS.497.3830F}.
In these filaments, the star formation continues
until the expanding H{\sc ii} regions completely disrupt them.

The spatial distributions of $\rm H_{2}$ and CO are also affected by the FUV feedback.
Once massive stars start to form, CO molecules can remain only in the high-density filaments (see Fig. \ref{fig_sigma100m4}-3).
Therefore, it is difficult to probe star-forming clouds only with CO molecule emissions, which are called "CO-dark" \citep{1988ApJ...334..771V, 2010ApJ...716.1191W}.
At $t\sim 2~{t_{\rm ff}}$, the filaments near the center are destroyed and
hydrogen molecules can distribute only in the outside high-density regions.

\begin{figure}
    \begin{center}
        \includegraphics[width=\columnwidth]{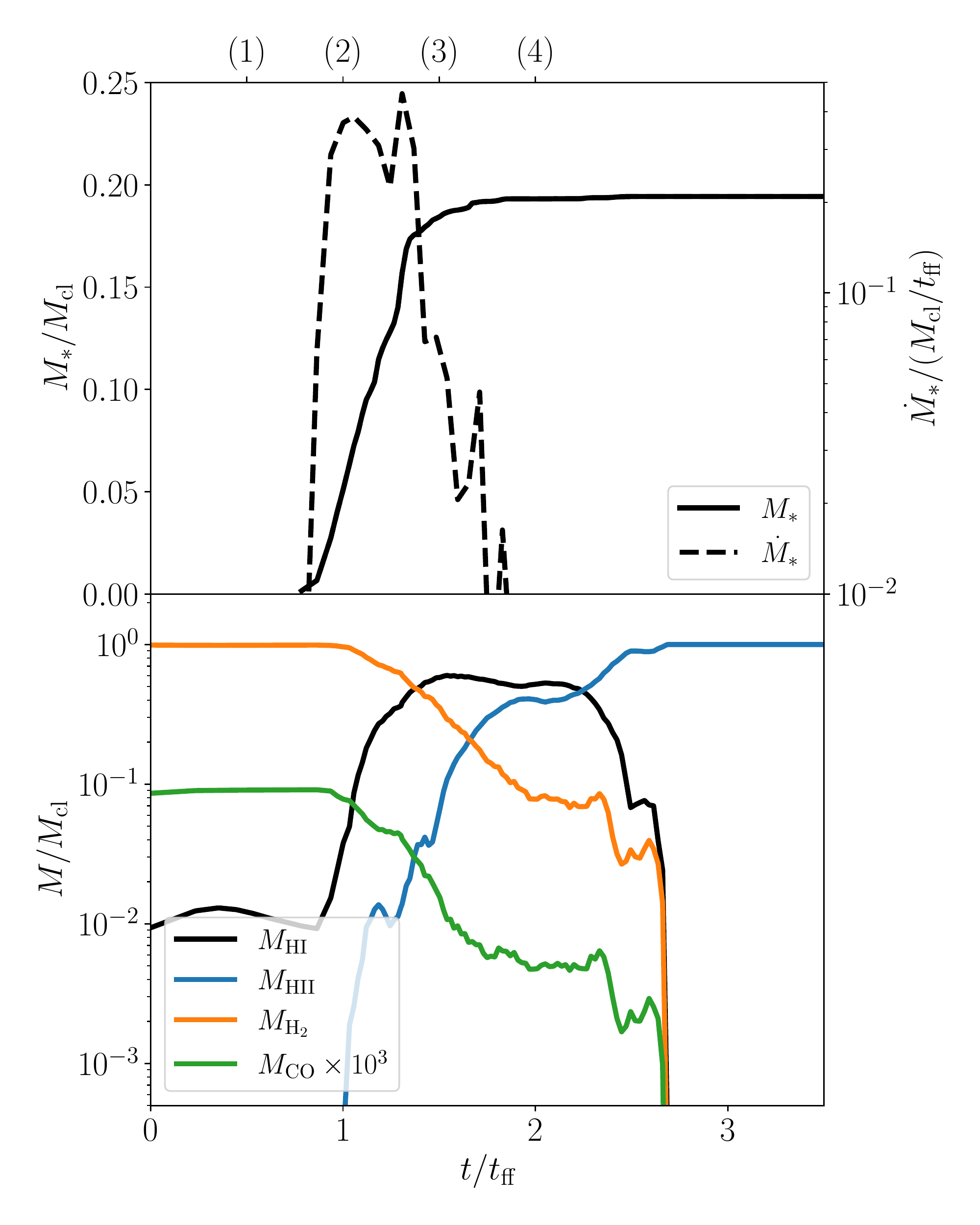}
    \end{center}
    \caption{Upper panel: The time evolution of stellar mass (solid) and star formation rate (dashed) normalized by the initial cloud mass ($M_{\rm cl}$) and $M_{\rm cl}/t_{\rm ff}$ where $t_{\rm ff}$ is the free-fall time.
    Lower panel: The masses of atomic hydrogen (H{\sc i}, black), ionized gas (H{\sc ii}, blue), hydrogen molecules ($\rm H_2$, orange), and CO molecules (CO, green).
    The amount of CO molecules is multiplied by $10^3$.
    The labels of (1)-(4) represents the four epochs as shown in Figure \ref{fig_sigma100m4}.}
    \label{fig_mevolv_s100m4EF}
\end{figure}

The abundances of chemical compositions change with the evolution of the cloud as shown in the bottom panel of Figure \ref{fig_mevolv_s100m4EF}.
At first, the gas is fully molecules until the star formation occurs ($t \sim t_{\rm ff}$).
Then, as stars form, hydrogen molecules are dissociated due to FUV feedback and converted into atomic hydrogen.
Thus, the abundances of molecules ($\rm H_2$ and CO) gradually decrease after $t\sim t_{\rm ff}$.
At $t \sim 1.3~t_{\rm ff}$, half of molecule gas is dissociated, and the SFR suddenly drops.
The abundance of ionized gas also starts to increases at this epoch.
At $t\sim 2~t_{\rm ff}$, the star formation is completely quenched.
However, as shown in Figure \ref{fig_sigma100m4}, the pillar structures remains, where $\sim 10$ percent of the cloud mass can survive as hydrogen molecules.
Finally, all gas is ionized and photoevaporated at $t\gtrsim 3~t_{\rm ff}$.

\subsubsection{EUV/FUV feedback effects}\label{Sec_EUV_FUV_feedback}

\begin{figure}
    \begin{center}
        \includegraphics[width=\columnwidth]{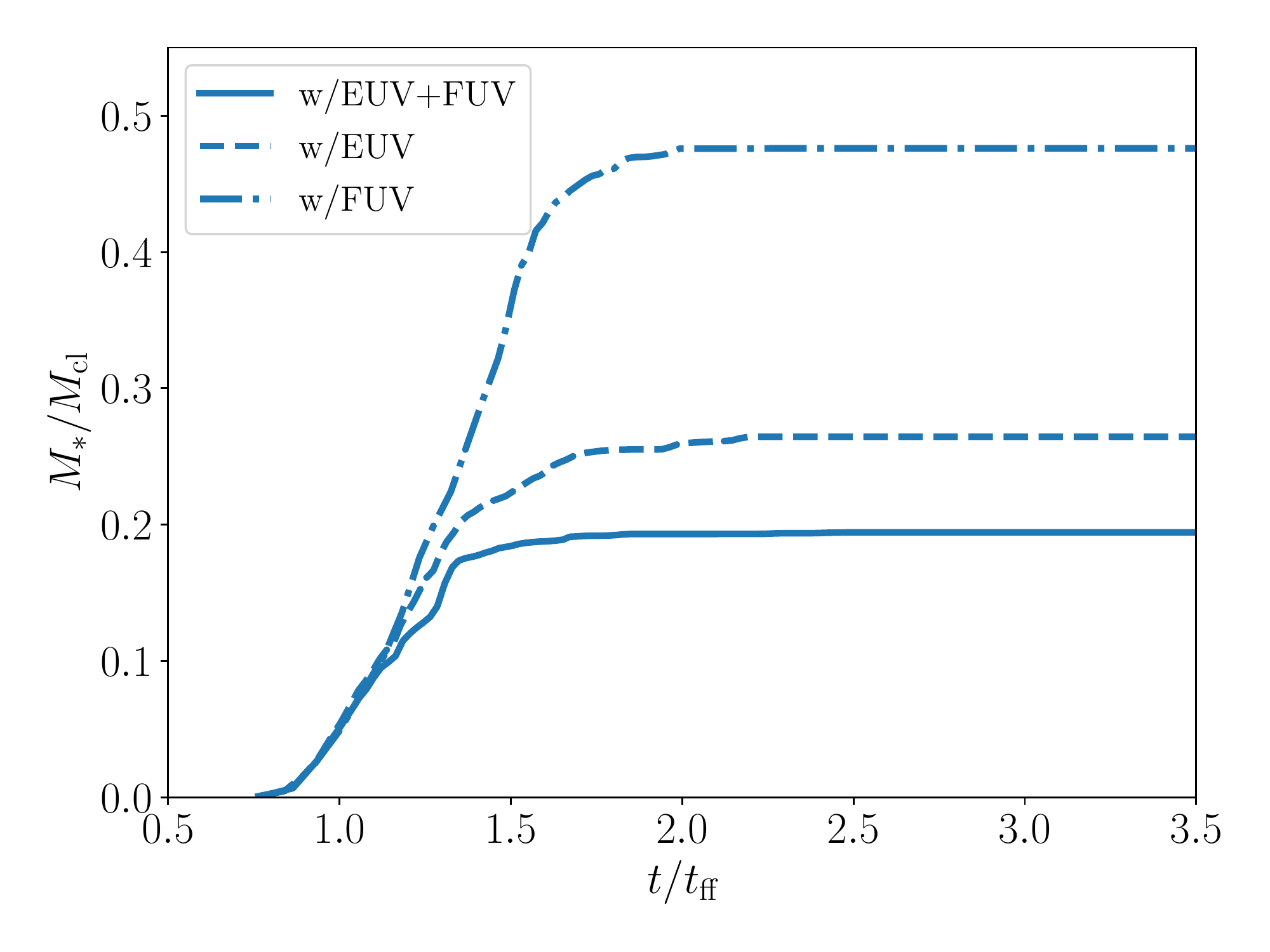}
    \end{center}
    \caption{The time evolution of stellar masses in the cases with $(\Sigma_{\rm cl}, M_{\rm cl})=(100~M_{\odot} {\rm pc^{-2}}, 10^4~M_{\odot})$.
    Each line represents the simulations with both EUV and FUV feedback (solid), only with the EUV (dashed) or FUV feedback (dot-dashed).}
    \label{fig:sig100m4}
\end{figure}

FUV photons can propagate beyond ionization fronts and heat up the large volume of the gas \citep[e.g.,][]{1998ApJ...501..192D, 2005ApJ...623..917H, 2006ApJ...646..240H, 2020MNRAS.497.5061I}.
However, previous studies of the FUV feedback have been limited only for spherical symmetric clouds.
Here, we perform the additional simulations only with EUV or FUV feedback alone to investigate the impacts of the FUV feedback in inhomogeneous density fields.
Hereafter, we label the model with both EUV and FUV feedback as EUV+FUV.

\begin{figure*}
\begin{center}
    \includegraphics[width=170mm]{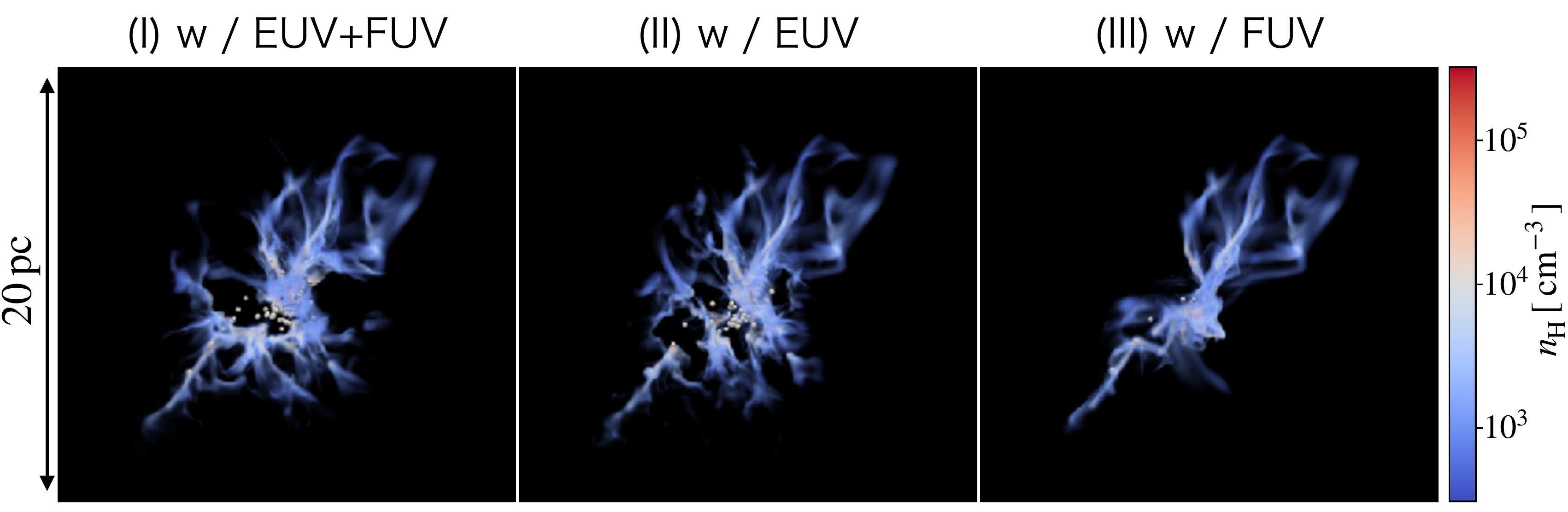}
\end{center}
    \caption{
    Gas structures visualized by the volume rendering technique in the cases with $(\Sigma_{\rm cl}, M_{\rm cl}) = (10^2~M_{\odot}{\rm pc^{-2}}, 10^4~M_{\odot})$.
    The white dots show the positions of sink particles.
    Each panel shows the different models: S100M4EF (left panel), S100M4E (middle panel), and S100M4F (right panel).
    The snapshots at $t = 3.3~{\rm Myr} (1.5~t_{\rm ff})$ are used.
    }
    \label{fig:nh_rendering}
\end{figure*}

Figure \ref{fig:sig100m4} shows the results in the cases with $(\Sigma_{\rm cl}, M_{\rm cl})=(100~M_{\odot} {\rm pc^{-2}}, 10^4~M_{\odot})$.
The star formation histories are almost the same until $t\sim 1~t_{\rm ff}$.
Then, the star formation rates rapidly increase only in the case with the FUV feedback alone.
We find that the star formation continues against the FUV feedback until the SFE becomes $\sim 0.4$.
Therefore, the FUV feedback alone cannot suppress the star formation significantly in this cloud case.
Figure \ref{fig:nh_rendering} shows the three-dimensional gas structures at the epoch of $t=1.5~t_{\rm ff}$.
In the model with the EUV feedback, the high-density regions are disrupted by the expanding H{\sc ii} bubbles.
Then, the cavities are created around the sink particles containing massive stars.
On the other hand, in the case only with FUV feedback, the high-density gas keeps distributing around the star cluster.
These results indicate that the EUV feedback is more effective in disrupting the dense gas than the FUV feedback.
Note that, however, the FUV feedback can enhance the strength of the EUV feedback.
The SFE with EUV+FUV is $\sim 20$ percent lower than the case only with EUV feedback.
This is because that the FUV feedback delays the star formation in part of dense gas until the ionization fronts propagate.

\begin{figure*}
    \begin{center}
    	\includegraphics[width=150mm]{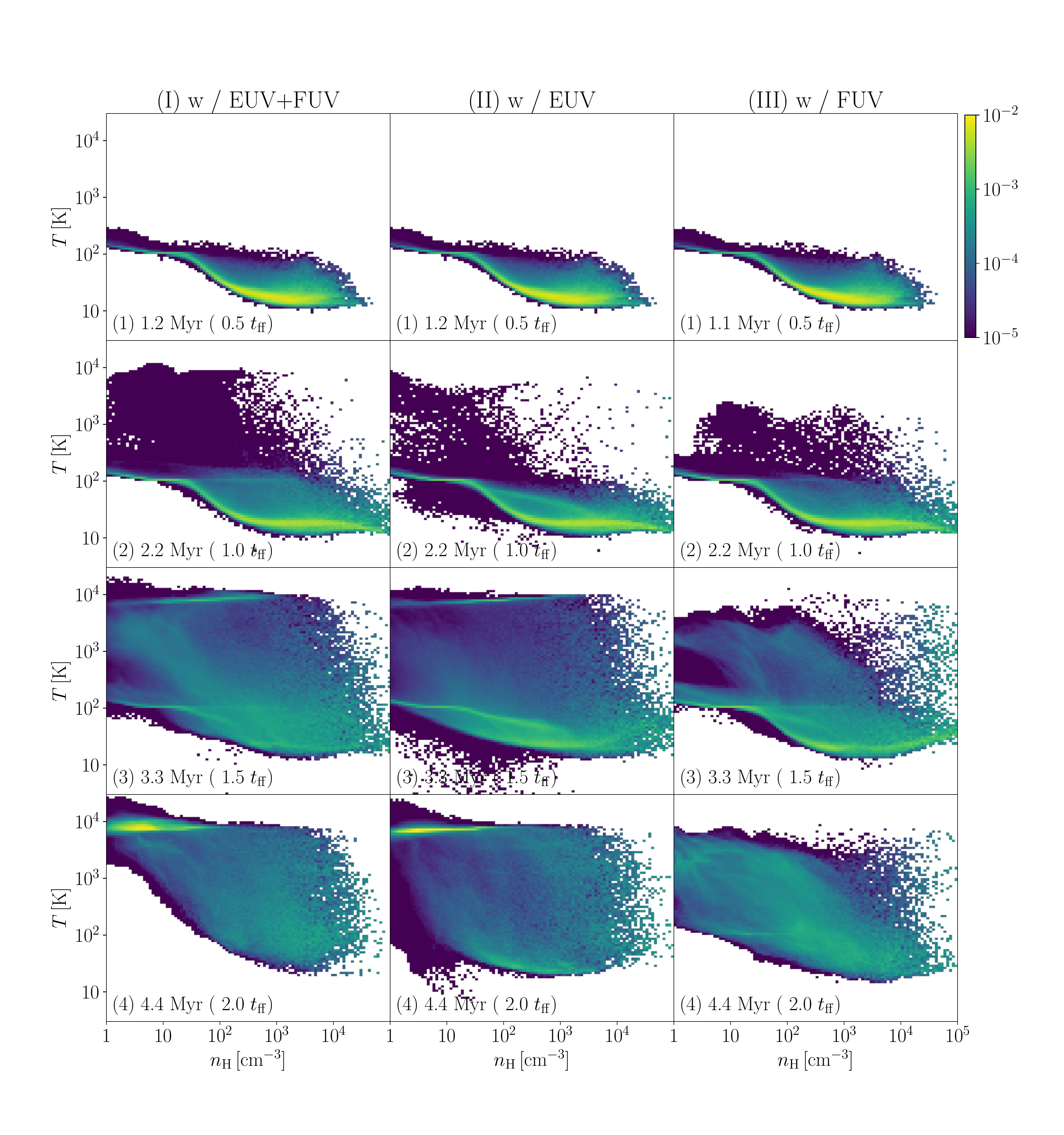}
    \end{center}
    \caption{
    The evolution of the physical conditions of the gas on the $n_{\rm H}-T$ plane in the models with $(\Sigma_{\rm cl}, M_{\rm cl}) = (10^2~M_{\odot}{\rm pc^{-2}}, 10^4~M_{\odot})$.
    The color contours are normalized by the total gas mass inside the simulation box.
    Each column show the included feedback effects; (I) the EUV and FUV feedback, (II) the EUV feedback, or (III) the FUV feedback.
}
    \label{fig_rhoT_sig100m4}
\end{figure*}

We investigate the heating effects of the EUV and FUV feedback.
Figure \ref{fig_rhoT_sig100m4} shows the evolution of the gas mass fractions in the $n_{\rm H} - T$ plane in the cases with $(\Sigma_{\rm cl}, M_{\rm cl}) =(10^2~M_{\odot}{\rm pc^{-2}}, 10^4~M_{\odot})$.
At $n_{\rm H} \sim 10^2-10^3~{\rm cm^{-3}}$, the gas temperature decreases to $\sim 10~{\rm K}$ due to the metal line cooling \citep[e.g.,][]{2005ApJ...626..627O, 2010ApJ...722.1793O}.
Turbulent motions can contribute to the heating of gas via shock.
At $n_{\rm H} \gtrsim 10^2~{\rm cm^{-3}}$, that is higher than the initial number density of the clouds, the gas is in the high temperature states \citep[e.g.,][]{2021arXiv210304997C}.
At $t\sim t_{\rm ff}$, stars start to form in the high-density gas, and their radiative feedback alters the distributions in the $n_{\rm H} - T$ plane.
In the model of EUV+FUV, the radiative feedback disrupts the clouds, and the star formation is almost quenched at $t\sim 1.5~t_{\rm ff}$.
At this epoch, the gas mass is widely distributed in the $n_{\rm H} - T$ plane depending on the place in the cloud.
The high-temperature states with $T \sim 10^4~{\rm K}$ represent the ionized regions.
At $t\sim 2 ~t_{\rm ff}$, most gas is ionized and distributes at $(n_{\rm H}, T) = (1-10^2~{\rm cm^{-3}}, \sim 10^4~{\rm K})$. A part of gas in the outside pillar-like structure remains as the cold neutral state (see Figure \ref{fig_sigma100m4}-4).

In the case with the EUV feedback alone, although the evolution is almost the same as that of EUV+FUV, there are a few different points.
At $t\sim 1.5~t_{\rm ff}$, more gas remains around the low-temperature regions at $n_{\rm H} \sim 10^1-10^3~{\rm cm^{-3}}$ and $T \sim 10^2 ~{\rm K}$.
These components correspond to the ambient gas outside H{\sc ii} regions, and FUV photons heat them in the model of EUV+FUV.
However, these low-density gas cannot convert to stars directly, not related to the SFEs.

In the model with the FUV feedback alone, most gas can keep the state of high-density and low-temperature even at $t\sim 1.5~t_{\rm ff}$.
Therefore, the star formation continues in these high-density gas.
This indicates that the FUV feedback cannot affect the star-forming gas directly.
The escape velocity of this cloud is $3.9~{\rm km/s}$ that corresponds to the sound speed of the gas with $T \sim 2.5 \times 10^3~{\rm K}$. Therefore, the gas in photodissociation regions (PDRs) is needed to be higher than this temperature to evaporate from the cloud.
Some low-density gas satisfies this condition and moves outward against the gravitational force of the cloud.
Whereas, the temperature of most high-density gas with $n_{\rm H} > 10^3~{\rm cm^{-3}}$ is lower than $T \sim 10^{3}~\rm K$, resulting in the longer duration time of the star formation and the higher SFE.

\begin{figure}
    \begin{center}
    	\includegraphics[width=\columnwidth]{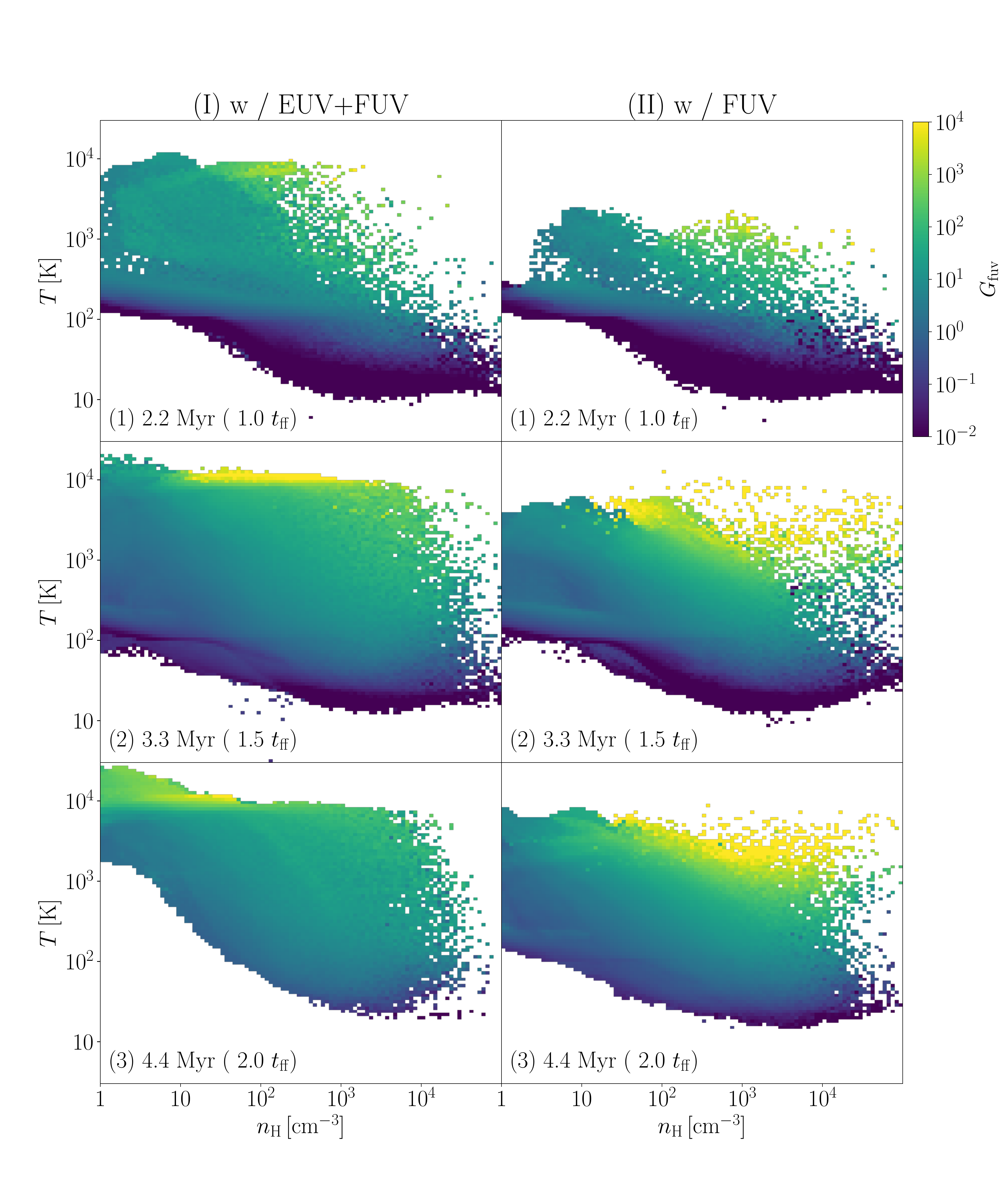}
    \end{center}
    \caption{
    The mean energy flux of FUV photons $G_{\rm fuv}$ in the $n_{\rm H}$-$\rho$ plane.
    The energy flux is normalized with the Habing flux.
    Here, $G_{\rm fuv} = 1$ corresponds to $1.6 \times 10^{-3} ~{\rm erg \, cm^{-2} \, s^{-1}}$.
    Same as Figure \ref{fig_rhoT_sig100m4}, each column shows the model with (I) the EUV and FUV feedback, or (II) the FUV feedback.
}
    \label{fig_rhoT_gfuv_sig100m4}
\end{figure}

In Figure \ref{fig_rhoT_gfuv_sig100m4}, we show the mean energy flux of FUV photons in the $n_{\rm H}$ - $T$ plane.
The flux is normalized by the Habing flux defined as
\begin{align}
    G_{\rm fuv} = \frac{F_{\rm FUV}}{F_{0}}, \label{eq_fuv_flux}
\end{align}
where $F_{\rm FUV}$ is energy flux at the FUV wavelength $6~{\rm eV} \leqq h\nu \leqq 13.6~{\rm eV}$, and $F_{0}$ is given as $F_0 = 1.6 \times 10^{-3} ~{\rm erg \, cm^{-2} \, s^{-1}}$.
In the model with EUV+FUV, the FUV energy flux exceeds $G_{\rm fuv} \gtrsim 10$ in the low-density regions of which the temperature becomes higher than $10^{3}~{\rm K}$ at $t\sim t_{\rm ff}$.
On the other hand, $G_{\rm fuv}$ is less than $1$ in the high-density gas ($n_{\rm H} \gtrsim 10^4~{\rm cm^{-3}}$).
This is because that the high-density filaments induced by the turbulent motions are optically thick for UV photons by dust attenuation \citep[e.g.,][]{2020MNRAS.497.3830F}. Also, even low-density gas can be shielded from the FUV radiation if it is behind the filaments, resulting in the low values of $G_{\rm fuv}$.
As the star formation proceeds, the filaments are destroyed and irradiated by the FUV radiaiton.
At $t\sim 2 ~t_{\rm ff}$, $G_{\rm fuv}$ gets higher in the entire regions as shown in Figure \ref{fig_rhoT_gfuv_sig100m4}-(I).

In the model with the FUV feedback, the gas temperature is mainly determined by the strength of the FUV radiation.
In particular, $G_{\rm fuv}$ higher than $10^4$ can heat the gas up to  $\gtrsim 10^3~{\rm K}$ at $n_{\rm H}> 10^3~{\rm cm^{-3}}$.
As discussed above, FUV feedback alone cannot heat the high-density gas inside the filaments.
These cold high-density components remain even at $t \sim 2~{t_{\rm ff}}$.
In a spherical uniform density cloud, once a massive star forms, the FUV feedback can quench the star formation rapidly via through the heating of a large volume of the gas \citep[e.g.,][]{2015A&A...580A..49I}.
Whereas, in the inhomogeneous density fields, the star formation can continue in the filaments even after massive stars form. Therefore, we suggest the impacts of the FUV feedback sensitively depend on the structure of the clouds.

\subsubsection{Dependence of SFEs on cloud masses and surface densities}\label{Sec_SFE_dependence_on_Mcl_Sig}

\begin{figure}
    \begin{center}
    	\includegraphics[width=\columnwidth]{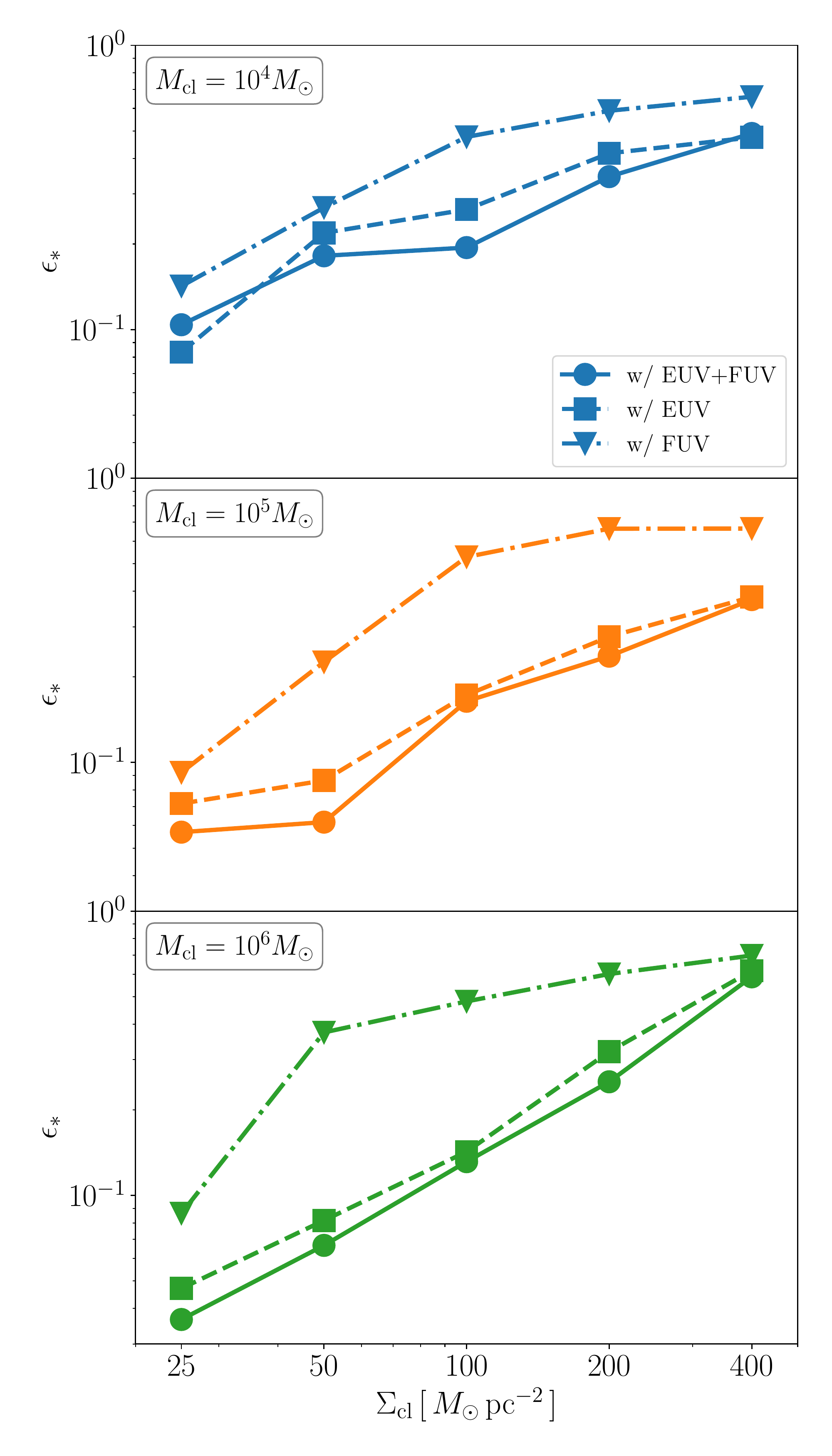}
    \end{center}
    \caption{
    The SFEs of clouds as a function of surface densities.
    Each panel shows the cases with $M_{\rm cl} = 10^4~M_{\odot}$, $10^5~M_{\odot}$, and $10^6~M_{\odot}$ from top to bottom.
    Each line shows the cases with both EUV and FUV feedback (solid), only with the EUV feedback (dashed), and the FUV feedback (dot-dashed).
}
    \label{fig_SFE}
\end{figure}

\begin{figure*}
    \begin{center}
    	\includegraphics[width=12cm]{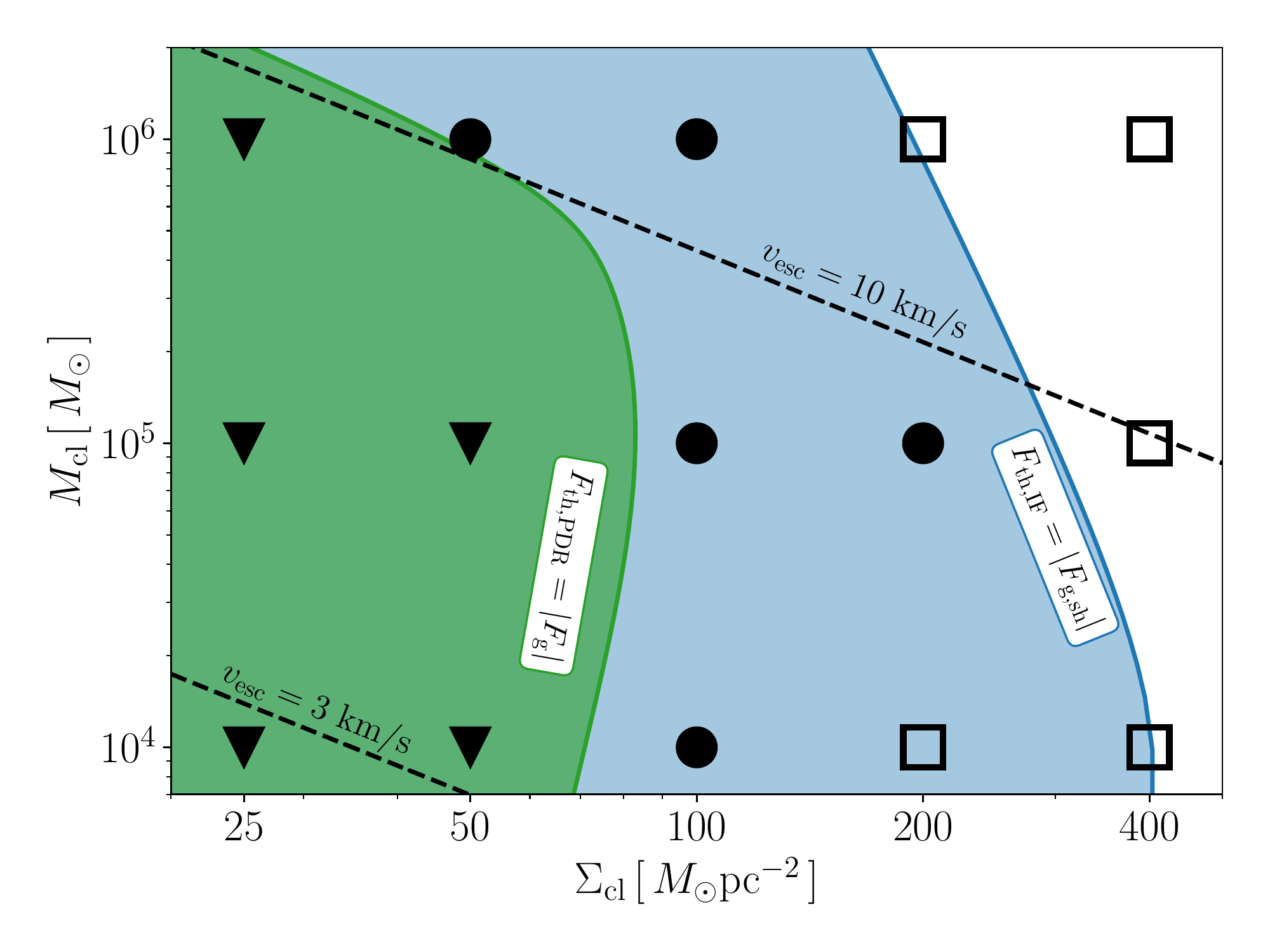}
    \end{center}
    \caption{
    Conditions for suppression of star formation by thermal pressure from H{\sc ii} regions $(F_{\rm th, IF})$ or PDRs $(F_{\rm th, PDF})$.
    The blue and green-shaded regions represent the conditions that thermal pressure from H{\sc ii} regions or PDRs exceeds gravitational forces from the clouds.
    The symbols indicate the effective feedback processes in the numerical simulations: (i) The FUV feedback alone can suppress the star formation and the SFEs become lower than 0.3 (triangles). (ii) The EUV feedback is required to suppress the star formation (circles). (iii) any radiative feedback is ineffective (squares).
}
    \label{fig_Pth_Pg}
\end{figure*}

The FUV feedback heats PDRs up to $\gtrsim 10^2~{\rm K}$, and it can disrupt the cloud if the cloud escape velocity is lower than the sound speed of PDRs.
Therefore, the impacts of the FUV feedback are likely to depend on the physical conditions of the clouds.
For example, in the cases with ($\Sigma_{\rm cl}$, $M_{\rm cl}$) = ($100~M_{\odot}{\rm pc^{-2}}, 10^{4}~M_{\odot}$), the SFE increases from 0.19 to 0.26 by turning off the FUV feedback.
Figure \ref{fig_SFE} shows the relations between SFEs and the initial surface densities of the clouds with $M_{\rm cl} = 10^4~M_{\odot}$, $10^5~M_{\odot}$, and $10^6~M_{\odot}$.
Previous studies suggested that the SFEs simply increase with the surface density given that both EUV and FUV feedback are included \citep[e.g.,][]{2010ApJ...710L.142F, 2016ApJ...829..130R, 2017MNRAS.471.4844G, 2018MNRAS.475.3511G, 2018ApJ...859...68K, 2019MNRAS.489.1880H, 2021MNRAS.506.5512F}.
In massive clouds with $M_{\rm cl} = 10^5~M_{\odot}$ and $10^6~M_{\rm \odot}$, the SFEs becomes  a few percent at $\Sigma_{\rm cl} \le 50~M_{\odot}{\rm pc^{-2}}$.
Whereas, in the cases of $M_{\rm cl} = 10^4~M_{\odot}$, the SFE is somewhat high $\sim 0.1$ even at the low surface densities.
It is related to the low probability of massive star formation if the total stellar mass is low.
Most stars have already formed before the EUV feedback becomes effective.
Then, the SFEs increase compared with the massive cloud cases.
We will further discuss the effects of the stochastic sampling on the radiative properties in Section \ref{Sec_stochastic_stellar_pop}.
We note that the SFEs of them are higher than the observed values in the local galaxies \citep[$\epsilon_* < 0.1$, e.g., ][]{2019Natur.569..519K, 2020MNRAS.493.2872C}.
In this study, we do not include the other feedback effects, especially outflow from the accreting protostars.
The outflow controls the SFEs in low-mass star formation \citep{2000ApJ...545..364M, 2013MNRAS.431.1719M}, and injects the momentum to the surrounding mediums \citep{2006ApJ...640L.187L, 2007ApJ...662..395N}.
In the low-mass clouds with $M_{\rm cl} \lesssim 10^4~M_{\odot}$, the outflow can mainly regulate the star formation before massive stars start to form \citep[see also,][]{2021MNRAS.502.3646G}.

In all models, the SFEs only with the FUV feedback are higher than the cases with the EUV feedback.
Thus, the EUV feedback has a dominant role in suppressing the star formation.
Only in the cases with $\Sigma_{\rm cl} \sim 25~M_{\odot}{\rm pc^{-2}}$, even the FUV feedback alone can suppress the star formation significantly because the thermal pressure in PDRs overcomes the gravitation force of the cloud.
As the clouds become massive and compact like ($\Sigma_{\rm cl}$, $M_{\rm cl}$) = ($400~M_{\odot}{\rm pc^{-2}}, 10^{6}~M_{\odot}$), even the EUV feedback is no longer effective, resulting in the high SFE \citep[e.g.,][]{2021MNRAS.506.5512F, 2020MNRAS.496L...1D, 2021MNRAS.tmp.2789D}.

\subsubsection{Analytical arguments}\label{Section_analytical_arguments}

As shown in section \ref{Sec_SFE_dependence_on_Mcl_Sig}, the FUV feedback alone suppresses the star formation in diffuse clouds.
In this section, we construct a semi-analytical model to understand the simulation results.

The FUV feedback can disrupt the clouds if the thermal pressure from PDRs is larger than the gravitational force.
Here, we compare them at the outer boundary of the cloud.
The gravitational force is given as
\begin{align}
    F_{\rm g} = - \frac{G M_{\rm cl}}{R_{\rm cl}^2}. \label{eq_grav}
\end{align}
The pressure gradient force is estimated as
\begin{align}
    F_{\rm th, PDR} = -\frac{1}{\rho} \frac{dP}{dr} \sim  \frac{c_{\rm s, PDR}^2}{R_{\rm cl}}, \label{eq_thmforce}
\end{align}
where $c_{\rm s, PDR}$ is the sound speed of the PDRs.
Assuming the energy equilibrium of heating and cooling processes as $\Lambda = \Gamma$
, we evaluate the temperature in PDRs.
The heating and cooling rates are given as
\begin{align}
    \Gamma = \Gamma_{\rm pe}, \label{eq_heating_process}
\end{align}
and
\begin{align}
    \Lambda = \Lambda_{\rm CII} + \Lambda_{\rm OI} + \Lambda_{\rm Ly\alpha} +\Lambda_{\rm rec}, \label{eq_cooling_process}
\end{align}
where $\Gamma_{\rm pe}$ is the photoelectric heating rate, $\Lambda_{\rm CII}$, $\Lambda_{\rm OI}$, $\Lambda_{\rm Ly\alpha}$ are C{\sc ii}, O{\sc i}, Ly $\alpha$ line cooling rates, and $\Lambda_{\rm rec}$ is the cooling rate by recombination between electron and grains.
The cooling rates of C{\sc ii} and O{\sc i} line are given by \citet{2003ApJ...587..278W}, and we use the Ly $\alpha$ cooling rate of \citet{1992ApJS...78..341C}.
We adapt the fitting formulae for the photoelectric heating and  the recombination cooling given in \citet{1994ApJ...427..822B}.
These rates are the functions of the energy flux of FUV radiation ($G_{\rm fuv}$) and the electron number density ($n_{\rm e}$).
The energy flux of FUV radiation is estimated as
\begin{align}
    F_{\rm FUV} = \frac{L_{\rm *, FUV}}{4 \pi R_{\rm cl}^2} = \frac{1}{4} \Psi_{\rm *, FUV} \epsilon_{*} \Sigma_{\rm cl}, \label{eq_energy_flux_fuv}
\end{align}
where $\epsilon_*$, $L_{\rm *, FUV}$ and $\Psi_{\rm *, FUV}$ are the SFE, luminosity and mass-to-luminosity ratio of FUV radiation.
Here, we use the relation of $L_{\rm *, FUV} = \Psi_{\rm *, FUV} \epsilon_{*} M_{\rm cl}$.
As shown in Figure \ref{fig_mass_rad}, the mass-to-luminosity ratio fluctuate when the total stellar mass is less than $10^4~M_{\odot}$.
Their median value decreases with the stellar mass.
Here, we use the fitting function of $\Psi_{\rm *, FUV}$ given as Equation \eqref{eq_psi_fuv}.
The electron number density is calculated by taking the balance between the recombination and the cosmic-ray ionization processes.
\begin{align}
    \xi_{\rm H} n_{\rm H} = \kappa_{\rm rec} n_{\rm H}^2 x_{\rm e}^2, \label{eq_rec}
\end{align}
where $\kappa_{\rm rec}$ is the recombination rate of ionized hydrogen and electron given in \citet{2003ApJ...587..278W}.
The ionization degree is calculated as
\begin{align}
    x_{\rm e} = 3.5 \times 10^{-4} \left( \frac{\xi_{\rm H}}{10^{-16} \, {\rm s^{-1}}} \right)^{1/2} \left( \frac{T}{10^2~{\rm K}} \right)^{3/8} \left( \frac{n_{\rm H}}{10^2 {\rm cm^{-3}}} \right)^{-1/2}. \label{eq_iondegree}
\end{align}

Figure \ref{fig_Pth_Pg} shows the analytical estimates with the simulation results.
The green-shaded region represents the clouds where the thermal pressure from the PDRs overcomes the gravitational force.
Here, we assume that the SFEs are $\epsilon_* = 0.3$.
In the blue-shaded region, the thermal pressure from H{\sc ii} regions ($F_{\rm th, IF}$) is larger than the gravitational force ($F_{\rm g, sh}$).
The derivation of $F_{\rm th, IF}$ and $F_{\rm g, sh}$ is shown in Appendix \ref{app_thermal_HII}.
In Figure \ref{fig_Pth_Pg}, each marker represents the effective feedback processes in the numerical simulations: (I) both FUV and EUV feedback ($\triangledown$), (II) the EUV feedback ($\bigcirc$), and (III) no radiative feedback is effective ($\square$).
Here, we set the threshold of the SFEs as $\epsilon_* = 0.3$ to determine whether each feedback is effective.
The analytical estimates reproduce well the simulation results.
The FUV feedback can suppress the star formation significantly only if the cloud surface density is lower than $\Sigma_{\rm cl} \sim 50~M_{\odot}{\rm pc^{-2}}$.
We note that the EUV feedback is also ineffective at $(\Sigma_{\rm cl}, M_{\rm cl}) = (200~M_{\odot}{\rm pc^{-2}}, 10^4~M_{\odot} )$ which is less than the threshold surface density predicted by the analytical estimates.
This is because low-mass stars form efficiently before the birth of massive stars due to the stochastic sampling manner.

\subsection{Dependence on virial parameter}\label{Sect_Dep_virial}
\begin{figure}
    \begin{center}
    	\includegraphics[width=\columnwidth]{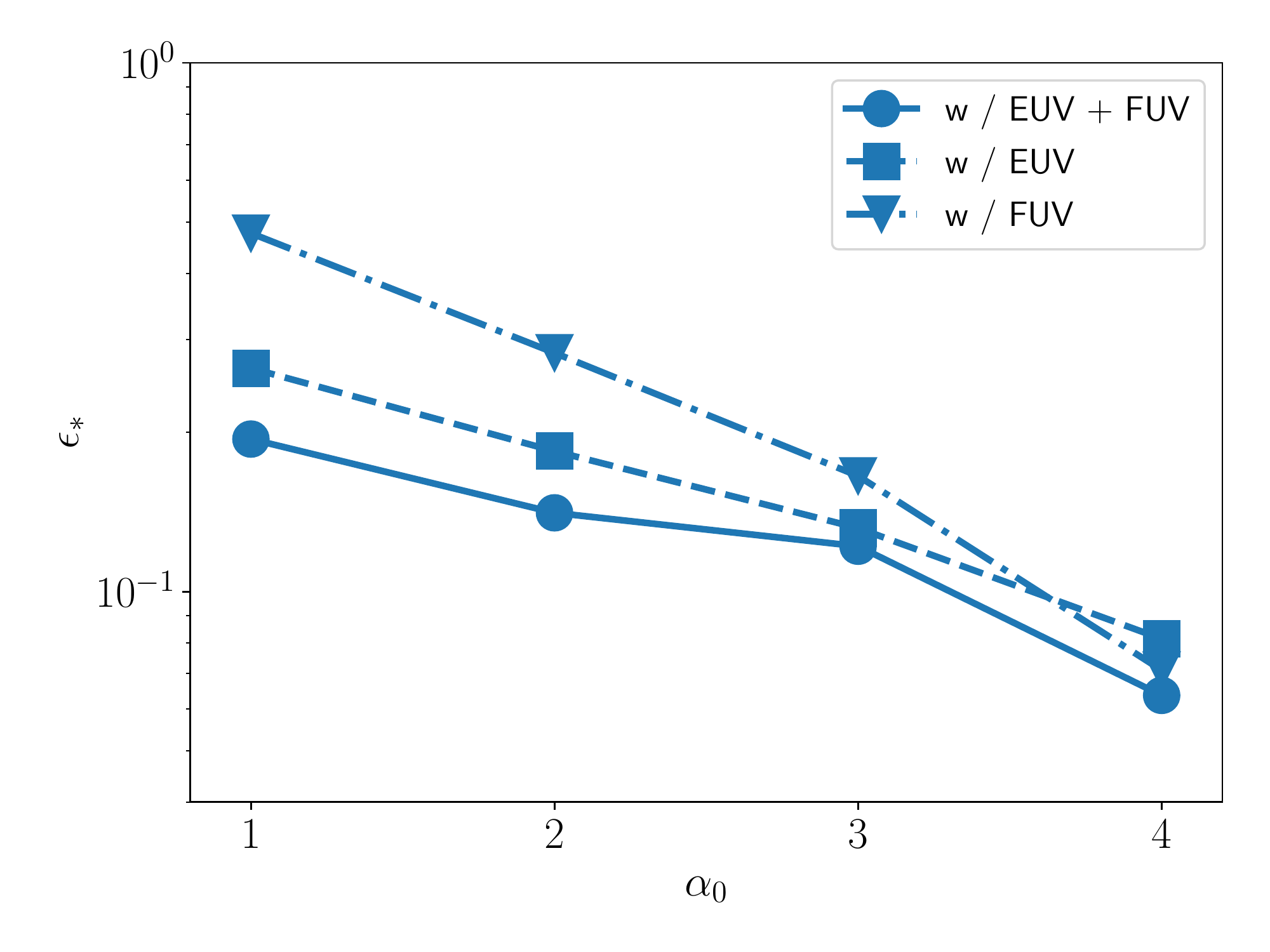}
    \end{center}
    \caption{
    The star formation efficiencies $(\epsilon_*)$ as a function of virial parameters $\alpha_0$ in the cases with $(M_{\rm cl}, \Sigma_{\rm cl}) = (10^4~M_{\odot}, 10^2~M_{\odot}{\rm pc^{-2}})$. Each line shows the cases with both EUV and FUV feedback (solid), only with the EUV (dashed), or the FUV feedback (dot-dashed).
}
    \label{fig_s100m4alpha}
\end{figure}

The collapse of clouds with high turbulence energy can proceed slowly.
In such a case, the star formation rate is likely to be lower.
Besides, the strong turbulent motions make low-density regions where
the radiative feedback efficiently suppresses the star formation \citep{2017MNRAS.467.1067D, 2021ApJ...911..128K, 2021MNRAS.506.5512F}.
Here, we study the impacts of the initial turbulence energy on the SFE.
Figure \ref{fig_s100m4alpha} shows the dependence of the SFEs on the virial parameters in the models with $(\Sigma_{\rm cl}, M_{\rm cl} )= (10^2~M_{\odot}{\rm pc^{-2}}, 10^4~M_{\odot})$.
The SFE decreases with higher virial parameters by a factor of 2-3.

In Figure \ref{fig_s100m4alpha}, we also show the results only with the EUV or FUV feedback.
As discussed in Section \ref{Sec_EUV_FUV_feedback}, the FUV feedback is the secondary effect in suppressing the SFEs at $\alpha_0=1$.
In the models with $\alpha_0 = 2$, the kinetic energy is comparable to the gravitational energy at the initial stages.
In our simulations, the turbulent motions decay efficiently, resulting in the rapid collapse of the cloud \citep[e.g.,][]{2011ApJ...740...74K}.
Thus, even for $\alpha_0 = 2$, the clouds are gravitationally bound when massive stars start to form.
In the cases with $\alpha_{0} \lesssim 2$, the EUV feedback plays a dominant role in regulating the SFE.
Whereas, in the cases with $\alpha_{0} \gtrsim 3$, even the FUV feedback alone can suppress the star formation significantly.
At $\alpha_0 = 4$, the SFEs are less than 0.1 in all cases.
These results are consistent with the simulation results of \citet{2017MNRAS.467.1067D}.
In the cases with $\alpha_{0} > 2$, the clouds are gravitationally unbound, and hence stars form only in the compressed high-density gas.
FUV photons propagate low-density regions induced by turbulence more rapidly than EUV ones and
heat the ambient gas before turbulent motions compress it.
Therefore, the FUV feedback is more efficient in suppressing the star formation in the unbound objects as predicted by \citet{1998ApJ...501..192D} and \cite{2015A&A...580A..49I}.

\begin{figure*}
    \begin{center}
    	\includegraphics[width=170mm]{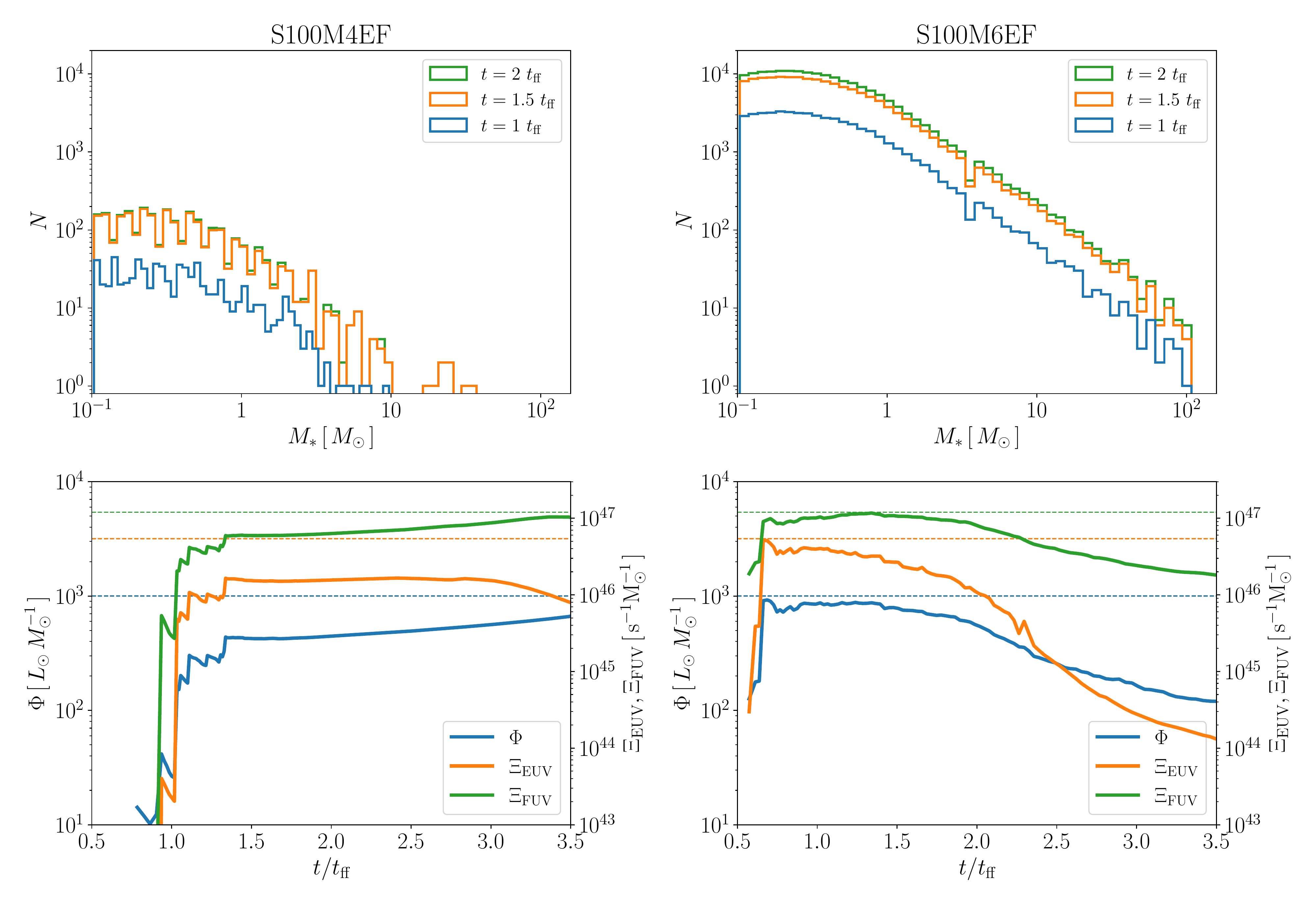}
    \end{center}
    \caption{
    Top panel: The stellar populations at $t=1$, $1.5$, and $2 ~t_{\rm ff}$. Bottom panel: Time evolution of the mass-to-luminosity ratio ($\Phi$), the emissivity of EUV and FUV photons per stellar mass ($\Xi_{\rm EUV}$ and $\Xi_{\rm FUV}$).
    Left panel: The model with ($\Sigma_{\rm cl}$, $M_{\rm cl}$) = ($100~M_{\odot}{\rm pc^{-2}}$, $10^4~M_{\odot}$). Right panel: The model with ($\Sigma_{\rm cl}$, $M_{\rm cl}$) = ($100~M_{\odot}{\rm pc^{-2}}$, $10^6~M_{\odot}$).
    The dashed lines represent the IMF-averaged emissivities.
    }
    \label{fig_lumistpop}
\end{figure*}

\subsection{Effects of stochastic stellar population on star cluster formation}\label{Sec_stochastic_stellar_pop}

\subsubsection{Radiative properties of star clusters}\label{Sec_rad_prop}

As pointed out in \citet{2016ApJ...819..137K}, a total stellar mass larger than $10^4~M_{\odot}$ is required to reproduce the modelled IMF in a stochastic manner.
If the total stellar mass is small, the stellar population is not fully sampled from the IMF.
This results in the deviation from the IMF-averaged value.
Top panels in Figure \ref{fig_lumistpop} show the mass distributions of the stellar components at $t=1$, $1.5$, and $2~t_{\rm ff}$ in the models of S100M4EF and S100M6EF.
In the clouds with $M_{\rm cl} = 10^4~M_{\odot}$, the total stellar mass is $2 \times 10^3~M_{\odot}$ at $t=2~t_{\rm ff}$.
Thus, the sampling effects appear.
At $t=t_{\rm ff}$, stars only with $< 10~M_{\odot}$ form.
Then, a few massive stars form until $t=1.5~t_{\rm ff}$ and quench the star formation via the radiative feedback.
Finally, they disrupt the host cloud.
The bottom panels in Figure \ref{fig_lumistpop} show the time evolution of the mass-to-luminosity ratio ($\Phi$), the emissivity of EUV and FUV photons per stellar mass ($\Xi_{\rm EUV}$ and $\Xi_{\rm FUV}$).
The emissivities are much smaller than the IMF averaged values until $t\sim t_{\rm ff}$, then increase rapidly as massive stars form. However, the emissivities are still lower than the IMF averaged values even after the quenching of the star formation.
On the other hand, in the case with $10^6~M_{\odot}$, the stellar populations reproduce the modelled IMF well.
In this model, massive stars can form in the early phase, and hence the emissivities are close to the IMF-averaged values.
At $t > 2~t_{\rm ff}$ ($t > 14.2~{\rm Myr}$), massive stars gradually reach the end of their lives.
Therefore, the emissivity decreases with the number of living massive stars.

\begin{figure}
    \begin{center}
    	\includegraphics[width=\columnwidth]{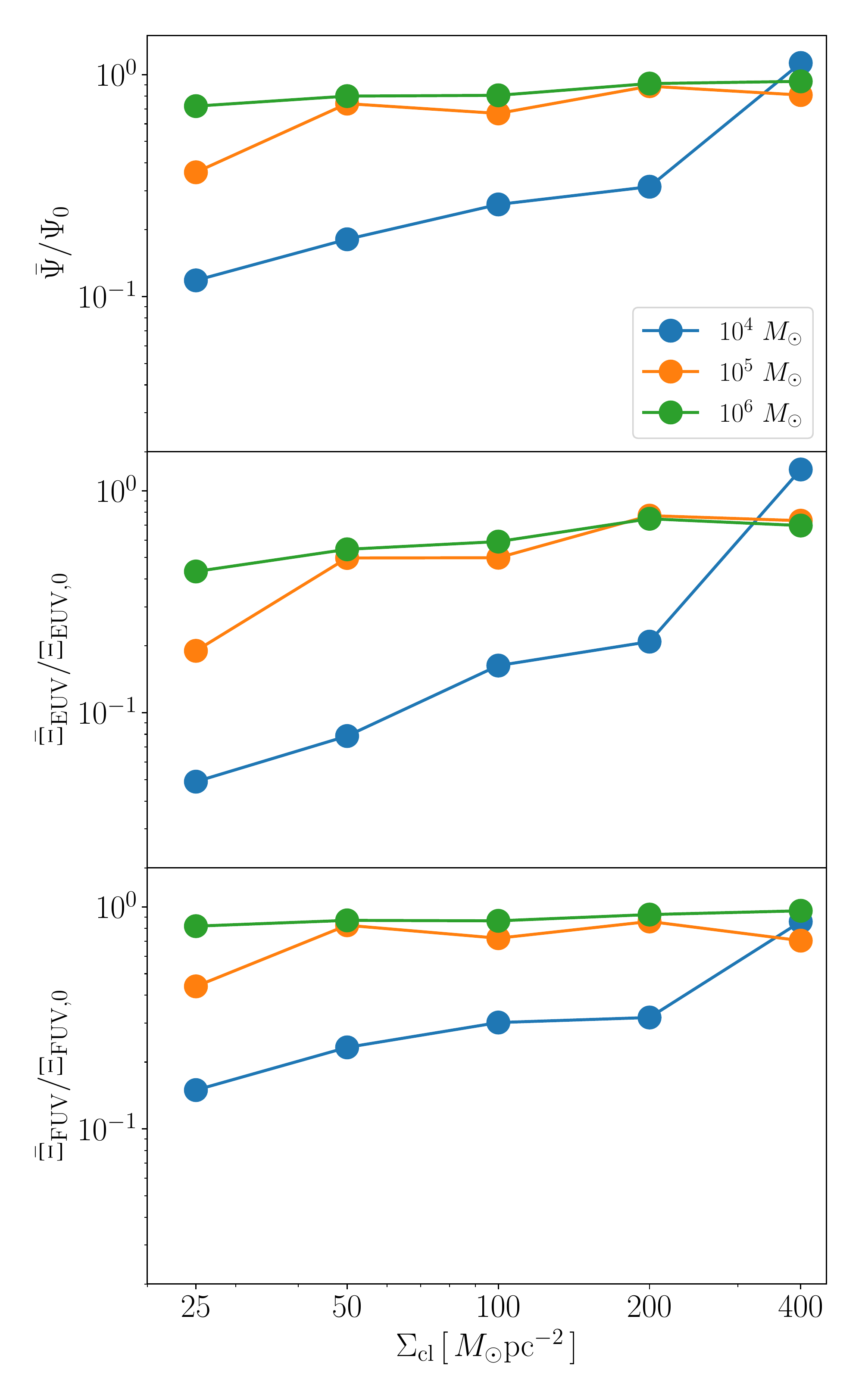}
    \end{center}
    \caption{
    Each panel shows the time-averaged light-to-mass ratios $(\bar{\Psi})$, and the emission rates of EUV and FUV photons $(\bar{\Xi}_{\rm EUV}, \bar{\Xi}_{\rm FUV})$ as a function of surface densities from top to bottom.
    They are normalized by the IMF-average values ($\Psi_0$, $\Xi_{\rm EUV, 0}$, and $\Xi_{\rm FUV, 0}$).
    Blue, orange and green lines represent the cases with $M_{\rm cl} = 10^4~M_{\odot}$ (blue), $10^5~M_{\odot}$ (orange), and $10^6~M_{\odot}$ (green).
}
    \label{fig_ltomratio}
\end{figure}

Figure \ref{fig_ltomratio} shows the time-averaged light-to-mass ratios ($\bar{\Psi}$) and the emission rates of EUV and FUV photons per stellar mass unit ($\bar{\Xi}_{\rm EUV}$ and $\bar{\Xi}_{\rm FUV}$) in each model.
We average the values from the onset of the star formation to the end of the cloud lives.
As shown in Figure \ref{fig_mass_rad}, the mass-to-light ratios tend to be lower than the IMF-averaged values if the total stellar mass is smaller than $\sim 10^{3}~M_{\odot}$ because the expected number of massive stars is below unity.
In low-mass clouds with $M_{\rm cl} = 10^4~M_{\odot}$, the emissivities increase with the surface densities.
In particular, the stellar mass becomes massive enough to reproduce the IMF at $\Sigma_{\rm cl} = 400~M_{\odot}{\rm pc^{-2}}$, resulting in the emissivity comparable to the IMF averaged value.
On the other hand, in the cases with the massive clouds of $10^{6}~M_{\odot}$,
the radiative properties do not change significantly and are close to the IMF averaged ones.

The lower emissivities than the IMF-averaged ones lead to higher SFEs as shown in Figure \ref{fig_SFE}.
In the low-mass clouds with $M_{\rm cl}=10^4~M_{\odot}$, it takes time for massive stars to form due to the low formation probability.
The radiative feedback is not strong and difficult to make the SFE lower than 0.2.
Therefore, the SFEs are higher than $\sim 0.1$ even for the low surface densities $\Sigma_{\rm cl} \lesssim 10^2~M_{\odot}$.
Whereas, in massive clouds with $M_{\rm cl}=10^5~M_{\odot}$ and $10^6~M_{\odot}$,
the SFEs simply decrease with surface densities as in the previous studies \citep[e.g.,][]{2010ApJ...710L.142F, 2018ApJ...859...68K, 2021MNRAS.506.5512F}.

\subsubsection{Properties of star cluster}\label{Sec_prop_star_cluster}

\begin{figure}
    \begin{center}
    	\includegraphics[width=\columnwidth]{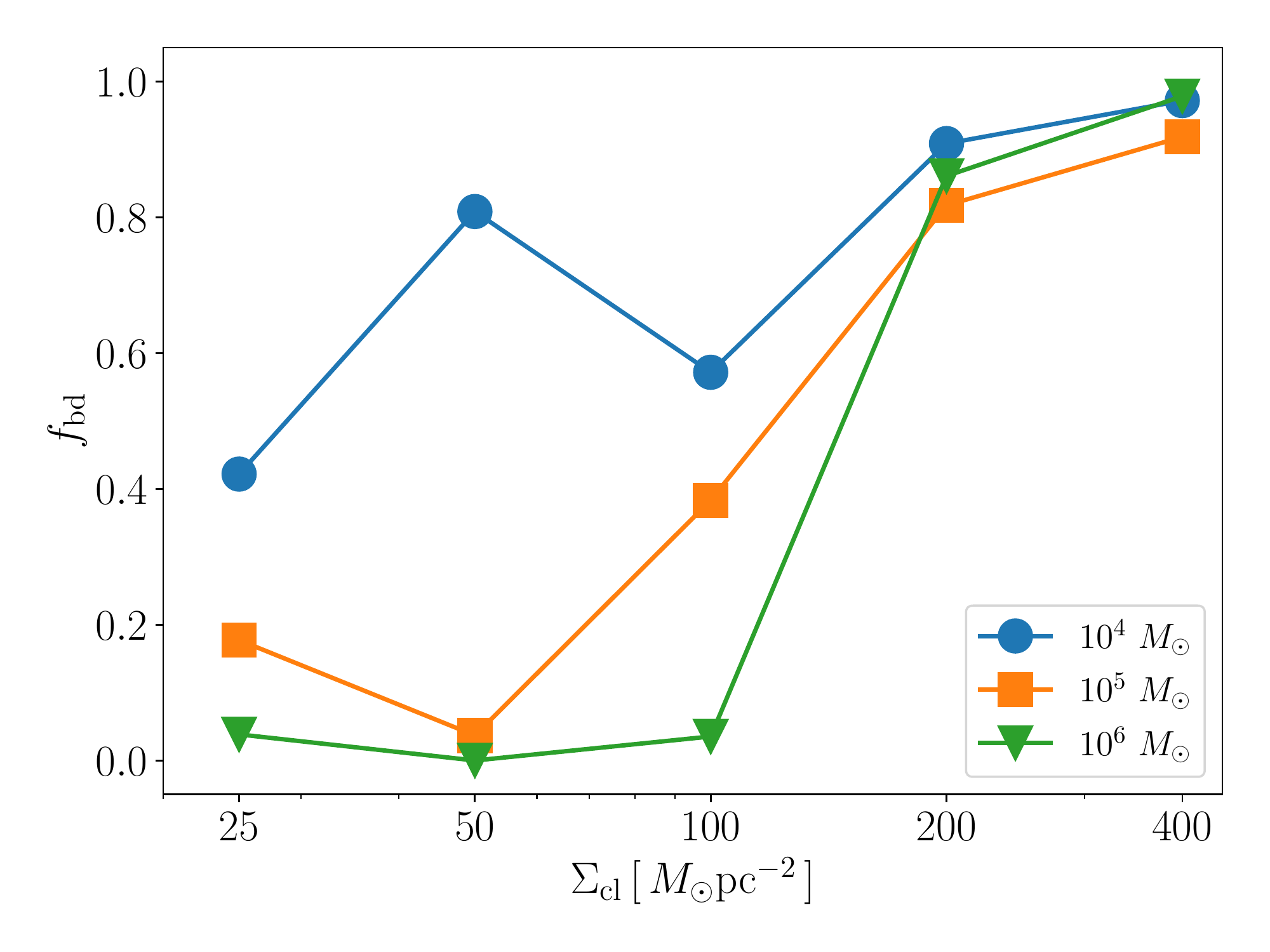}
    \end{center}
    \caption{
    The bound fractions at the end of the star formation as a function of surface densities.
    Blue, orange and green lines show the cases with $M_{\rm cl} = 10^4~M_{\odot}$ (blue), $10^5~M_{\odot}$ (orange), and $10^6~M_{\odot}$ (green).
}
    \label{fig_bdfrac}
\end{figure}

We here study the evolution of the star clusters after the cloud disruption by introducing the bound fraction as $f_{\rm bd} = M_{\rm bd}/ M_*$ where $M_{\rm bd}$ is the stellar mass gravitationally bound.
The bound fraction is related to the SFE in the star-cluster formation \citep[e.g.,][]{2000ApJ...542..964A, 2007MNRAS.380.1589B, 2017A&A...605A.119S, 2019MNRAS.487..364L, 2021MNRAS.506.3239G}.
If the SFE is high enough, the clusters keep gravitationally bound.
\citet{2007MNRAS.380.1589B} performed the N-body simulations of star clusters incorporating gas removal from clouds.
They showed that the star clusters are gravitationally bound even after the instantaneous removal of gas if the SFEs are higher than 0.33.
The SFE threshold can be lower if the cloud disruption proceeds slowly.
In \citetalias{2021MNRAS.506.5512F}, we showed that the bound fraction rapidly increases when the SFE exceeds $\sim 0.1-0.2$.
Figure \ref{fig_bdfrac} shows the bound fractions as a function of $\Sigma_{\rm cl}$.
For the clouds with $10^4~M_{\odot}$, the bound fractions exceed 0.6 even at $\Sigma \sim 50~M_{\odot}{\rm pc^{-2}}$.
In these models, the SFEs exceed the threshold value $\sim 0.1$.
Besides, the gas evaporation proceeds slowly in these cases because of the low formation probability of massive stars. This time period allows the cooling and condensation of the gas, resulting in a deeper gravitational potential well and the high $f_{\rm bd}$.
On the other hand, $f_{\rm bd}$ decreases in the cases with $M_{\rm cl} = 10^5~M_{\odot}$ and $10^6~M_{\odot}$.
As shown in Figure \ref{fig_ltomratio}, the emissivity of star clusters in the massive clouds is similar to the IMF-averaged value.
Therefore, massive stars can form earlier and disrupt the cloud rapidly. In such a situation, the star clusters can be virialized with the gravitational potential of the gas well, and become unbound after the cloud disruption.
At $\Sigma_{\rm cl} \gtrsim 200~M_{\odot}{\rm pc^{-2}}$, the SFEs exceed 0.2, and the bound fractions also increases to $>0.8$.
This rapid increase of the bound fraction at $\epsilon_* > 0.2$ is consistent with the previous works.

\begin{figure*}
    \begin{center}
    	\includegraphics[width=12cm]{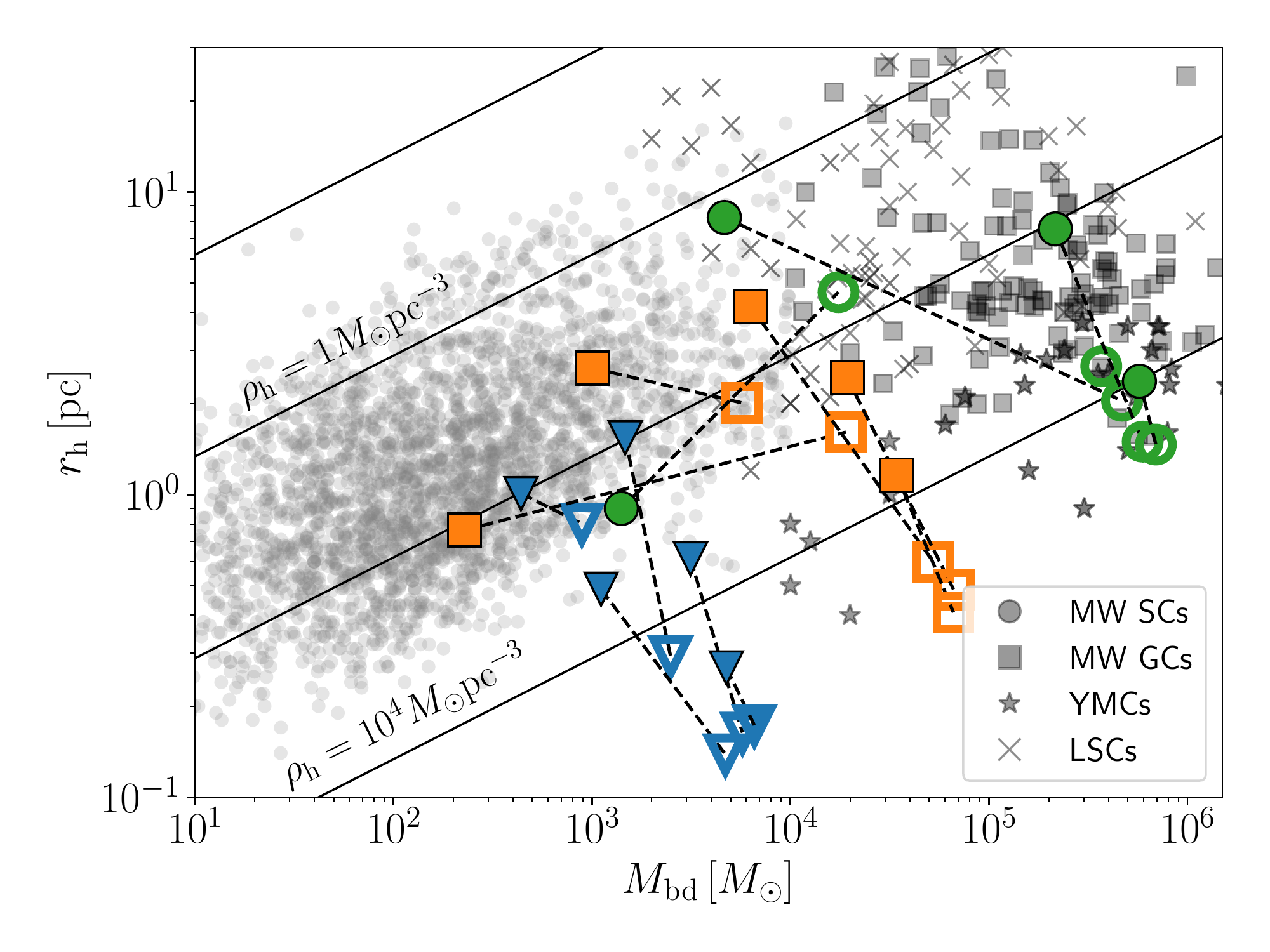}
    \end{center}
    \caption{The half-mass radius and mass relation of the star clusters obtained from the simulations..
    Each symbol shows the cases with $M_{\odot} = 10^4~M_{\odot}$ (blue), $10^5~M_{\odot}$ (orange), and $10^6~M_{\odot}$ (green).
    The filled  circles represent the results considering both EUV and FUV feedback, while the open ones show cases only with the FUV feedback.}
    The solid lines represent the stellar number densities $\rho_{\rm h}=10^{-2}$, $1$, $10^2$, and $10^4~M_{\odot}{\rm pc^{-2}}$ from top to bottom.
    The black markers represent the Milky Way neighborhood  star clusters (circle, MW SCs), globular clusters (square, MW GCs), young massive star clusters (star, YMCs), and leaky star clusters (cross, LCs).
    The lists of star clusters taken from \citet{2013A&A...558A..53K} (MW SCs), \citet{2018MNRAS.478.1520B} (MW GCs), and \citet{2010ARA&A..48..431P} (YMCs and LSCs).
    We set the threshold between YMCs and LCs at the stellar density $\rho_* = 10^3~M_{\odot} {\rm pc^{-3}}$ defined by \citet{2010ARA&A..48..431P}.
    \label{fig_Mbd}
\end{figure*}

Figure \ref{fig_Mbd} shows the relation between the half-mass radius and the stellar mass gravitationally bound.
In the clouds with $M_{\rm cl}= 10^4~M_{\odot}$, the bound objects remain regardless of surface densities of the clouds.
The masses of these star clusters are $M_* \geq 10^2~M_{\odot}$.
The stellar densities are comparable to the observed young star clusters \citep[e.g.,][]{2003ARA&A..41...57L}, and the results in the previous works \citep[e.g.,][]{2015PASJ...67...59F, 2016ApJ...817....4F}.
The masses and stellar densities of star clusters in the clouds with $M_{\rm cl} = 10^4~M_{\odot}$ are similar to the observed open clusters in the Milky Way.

In recent observations, the massive and low-density clusters are found ($\gtrsim 10^4 ~M_{\odot}$ and $1-10~M_{\odot} {\rm pc^{-2}}$), which are dubbed as leaky star clusters \citep{2009A&A...498L..37P, 2011A&A...536A..90P, 2016ApJ...817....4F}.
As mentioned in \citetalias{2021MNRAS.506.5512F}, the leaky star clusters are formed in low-surface density clouds where the stellar core formation does not occur.
In the massive clouds with $M_{\rm cl} \gtrsim 10^5~M_{\odot}$, the bound star clusters cannot remain or low-mass ones are formed when the surface densities are less than $\Sigma_{\rm cl} \sim 200~M_{\odot}{\rm pc^{-2}}$.
These star clusters remained as bound objects are similar to the leaky star cluster or low-mass star clusters ($M_* < 10^4~M_{\odot}$).
On the other hand, the SFEs and bound fractions are enhanced in the compact clouds with $\Sigma_{\rm cl} \gtrsim 200~M_{\odot}{\rm pc^{-2}}$.
In these cases, massive and high-dense star clusters are formed.
These star clusters are categorized as young massive star clusters \citep{2010ARA&A..48..431P}.

We find that the stellar densities become much higher than $\sim 10^4~M_{\odot}\; \rm pc^{-2}$ for the clouds with $M_{\rm cl}=10^{4}~M_{\odot}$ if the EUV feedback is turned off. These cases remain the compact star clusters with the masses of $\sim 10^{3} - 10^{4}~M_{\odot}$ that are unreasonable compared to the observed star clusters. Therefore, the FUV feedback alone is not enough to form low-mass stars with a reasonable size.
The EUV feedback induces expanding H{\sc ii} bubbles that prevent the gas inflow onto the center of the cloud, and the compressed gas shells can form next stars \citep{2021MNRAS.506.5512F}.
In particular, for diffuse clouds, the difference of the stellar densities with and without the EUV feedback becomes significant.
Note that, if both FUV and EUV feedback are turned off, most gas solely accretes onto a few sink particles at the center of the cloud and the stellar density becomes too high.
Thus, we suggest that the observed variety of star clusters is tightly related to the physical conditions of initial gas clouds and the radiative feedback.

\section{Discussion}\label{sec_discussion}
In this study, we adopt the gravitationally bound clouds as the initial conditions.
However, massive clouds can form gradually via the accumulation of gas.
Massive clouds in the Galactic disk always host stars \citep[e.g.,][]{2012ApJ...758L..29G}, while a few starless clouds are observed in the Galactic center \citep{2013MNRAS.433L..15L, 2014prpl.conf..291L}.
To explain the lack of starless clouds in the Galactic disk, \citet{2014prpl.conf..291L} proposed a "conveyor belt" model in which the interstellar gas accretes onto the cloud, allowing new star formation episodes.
Besides, \citet{2020MNRAS.494..624K} suggested that the conveyor belt model could explain the age distributions in the Galactic open clusters naturally.
\citet{2019MNRAS.490.3061V} showed the "global hierarchical collapse" (GHC) scenario in which there is a time delay between collapses of small and large scales in a cloud.
The GHC causes gas flow from the large scale into the local potential minimum, which corresponds to the "conveyor belt" model.
\citet{2020A&A...642A..87K} found that all observed O-type stars form at hubs as the crossing places of gas filaments.
In their scenario, the low-mass stars form along the filaments in a long time ($\sim 1 ~{\rm Myr}$), but massive star formation occurs in the short timescales ($\sim 0.1~{\rm Myr}$) at the hubs of the hub filaments.
As shown in our study, FUV radiation propagates outward rapidly in low-mass clouds, which is likely to affect the dynamics of the unbound gas, such as the accretion gas or filaments.
By performing RHD simulations with larger calculation boxes, we will develop a unified model of the cloud formation with the star cluster formation and investigate the above scenarios and the impacts of the FUV feedback on the global structure in future studies.

The observational studies have suggested that massive stars tend to form only in high-density environments \citep{2021PASJ...73S...1F}.
\citet{2010ApJ...723L...7K} showed that clouds need to satisfy $M_{\rm cl}\gtrsim 870~M_{\odot} (R_{\rm cl}/{\rm pc})^{1.33}$ for the massive star formation \citep[see also,][]{2018MNRAS.473.1059U, 2021PASJ...73S..75E}.
Diffuse clouds in this work do not satisfy this condition.
Nevertheless, our simulations allow the diffuse clouds to form massive stars.
Besides, the SFEs are typically 0.2 in these clouds, and it is higher than the observed values in the local galaxies \citep{2019Natur.569..519K, 2020MNRAS.493.2872C}.
In diffuse clouds without massive stars, the bipolar outflow launched from low-mass stars becomes the main feedback effect \citep[e.g.,][]{2000ApJ...545..364M, 2006ApJ...640L.187L, 2007ApJ...662..395N}.
Recently, \citet{2021MNRAS.502.3646G} found that the low-surface density cloud $\Sigma_{\rm cl} \sim 60~M_{\odot}{\rm pc^{-2}}$ with the mass of $2 \times 10^4~M_{\odot}$ is disrupted by the outflow effect,  and it regulates the star formation.
Thus, in cases with low-surface densities, the outflow can be a key to understanding the star formation, which is out of scope in the current studies.

In addition to the outflow from low-mass stars, SNe and stellar wind are likely to affect the cloud dynamics and the star formation.
In some cases of our simulations, the duration time of the star formation is longer than the lifetimes of OB stars.
In these clouds, SNe occurs and evacuates the gas from the clouds \citep[e.g.,][]{2016MNRAS.463.3129G}.
Stellar wind pushes out surrounding gas \citep[e.g.,][]{2014MNRAS.442..694D, 2020MNRAS.497.4718D, 2021MNRAS.501.1352G, 2021arXiv210712397R}, and heats up the gas to $(\gtrsim 10^5~{\rm K})$ via the shock \citep[e.g.,][]{2021ApJ...914...89L, 2021ApJ...914...90L}.
Indeed, X-ray emission is observed from the region inside the expanding shell \citep{2021SciA....7.9511L}, which is likely to be induced by the stellar wind.
We will include these effects in future works.

In our simulations, we have calculated non-equilibrium chemical reactions of the species, such as CO and $\rm H_2$ molecules, and C{\sc ii}.
The spatial distributions of these abundances can be tools to probe the different evolutional stages of the star cluster formation \citep[e.g.,][]{2021arXiv210809018T}. For example, the cavity structures are formed when the expanding H{\sc ii} bubbles trigger the star formation in clouds with $M_{\rm cl} \lesssim 10^4~M_{\odot}$.
In this case, the CO molecules only remain inside the high-density shell against the photodissociation feedback, and thus the arch structure is observed with CO line emission.
We plan to investigate the trigger mechanism of the star formation by modeling the maps of metal line emissions, such as CO and C{\sc ii}, and then compare with the observational results in future study.

\section{Summary}\label{Section_Summary_and_Discussion}

We have performed the 3D RHD simulations with the stochastic stellar population models.
Our simulations include radiative feedback from massive stars, such as photoionization, photodissociation of molecules, and photoelectric heating.
We have investigated the star cluster formation in clouds with various masses and surface densities, $M_{\rm cl} = 10^4-10^6~M_{\odot}$ and $\Sigma_{\rm cl} = 50-400~M_{\odot}{\rm pc^{-2}}$.
We have also studied the effects of the EUV and FUV feedback on the cloud disruption process.
Our findings are summarized as follows:

\begin{itemize}
    \item[(i)]
     For diffuse clouds with $\Sigma_{\rm cl} \lesssim 25-50~M_{\odot}{\rm pc^{-2}}$,
     the FUV feedback alone can regulate the SFEs to be less than 30 percent.
     In these cases, FUV radiation rapidly propagates in a cloud and heats up the gas via the photoelectric heating and dissociation of hydrogen molecules.
     The thermal pressure from the warm PDRs disrupt the clouds and quench the star formation. The clouds with the mass of $10^{4}-10^{5}~M_{\odot}$ and the low surface densities are likely to form low-mass star clusters as observed open clusters in the Milky Way.

     \item[(ii)]
     The EUV feedback mainly suppresses the star formation in the clouds with $\Sigma_{\rm cl} = 100-200 ~M_{\odot}{\rm pc^{-2}}$. Massive stars create H{\sc ii} bubbles that expand rapidly due to the high thermal pressure.
     For clouds with the masses of $10^{4}-10^{5}~M_{\odot}$, star clusters become too compact compared with observed ones if the EUV feedback is not taken into account. Therefore, the EUV feedback plays an essential role in the formation of  star clusters with reasonable mass and size.

     \item[(iii)] Once the initial surface density exceeds $200 ~M_{\odot}{\rm pc^{-2}}$, even thermal pressure from H{\sc ii} regions cannot evacuate the gas against the gravitation force from the clouds. In such a case, high-density stellar cores form, and the SFEs exceed 0.3.
     In particular, young massive star clusters are formed in the massive compact clouds with $M_{\rm cl} \gtrsim 10^5~M_{\odot}$.

  \item[(iv)]
     In low-mass clouds with $M_{\rm cl} = 10^4~M_{\odot}$, the number of stars is not enough to reproduce the modelled IMF smoothly.
     Therefore, a part of gas can be converted into stars before massive stars start to form, resulting in the SFE higher than $\sim 0.1$ even for diffuse clouds with $\Sigma_{\rm cl} \lesssim 200~M_{\odot}$.

    \item[(v)] Turbulent motions delay the star formation, and thus the SFEs decrease with higher virial parameters. According to the simulations with the different virial parameters for clouds with $(\Sigma_{\rm cl}, M_{\rm cl}) = (10^2~M_{\odot}{\rm pc^{-2}}, 10^4~M_{\odot})$, the SFEs gradually decreases from 0.19 to 0.06 in the rages of $\alpha_0 = 1-4$. At $\alpha_{0}\lesssim 3$,
    the star formation is regulated mainly by the EUV feedback. In the unbound clouds with $\alpha_0=4$, FUV feedback hampers the gravitational collapse of clumps and reduces the SFEs.

\end{itemize}

FUV photons create PDRs where rapidly propagate beyond the H{\sc ii} regions. Consequently, a large volume of gas becomes the warm H{\sc i} state with the temperature $\gtrsim 10^2~{\rm K}$. Therefore, the FUV feedback has been considered to reduce the SFEs in the GMCs  \citep[e.g.,][]{1998ApJ...501..192D, 2015A&A...580A..49I}.
However, the impacts of the FUV feedback can be secondary in cases with inhomogeneous filamentary structures.
The star formation mainly occurs in the high-density filaments where the FUV feedback cannot disrupt these structures due to the dust absorption of FUV photons.
Hence, the shielding effects of these filaments weaken the heating effects on the ambient low-density gas.
Besides, the thermal pressure from PDRs cannot evacuate the gas if the escape velocity of clouds is much larger than the sound speeds in PDRs.
On the other hand, the FUV feedback is more effective if clouds are gravitationally unbound due to strong turbulent motions.
Therefore, the physical states of clouds alter the role of the FUV feedback in star cluster formation.

Thus, we have shown that the populations of formed star clusters sensitively depend on the initial conditions of clouds and the radiative feedback.
We will investigate the star cluster formation with the cloud formation consistently in future work.

\section*{Acknowledgements}
The authors wish to express their cordial thanks to Profs. Masayuki Umemura and Ken Ohsuga for their continual interest, advice, and encouragement.
We appreciate Tomoaki Matsumoto for great
contribution to the code development.
We would like to thank Takashi Hosokawa, and Shu-ichiro Inutsuka for useful discussions and comments.
The numerical simulations were performed on the Cray XC50 (Aterui II) at the Center for Computational Astrophysics of National Astronomical Observatory of Japan and Yukawa-21 at Yukawa Institute for Theoretical Physics in Kyoto University.
This work is supported in part by MEXT/JSPS KAKENHI Grant Number 17H04827, 18H04570, 20H04724, 21H04489 (HY), NAOJ ALMA Scientific Research Grant Numbers 2019-11A  (HY), and JST FOREST Program, Grant Number JP-MJFR202Z (HY).

\section*{Data Availability}

The data underlying this article will be shared on reasonable request to the corresponding author.



\bibliographystyle{mnras}
\input{main.bbl}




\appendix
\section{Dependence on random seeds}\label{Sec_random_seed}

\begin{figure*}
    \begin{center}
    	\includegraphics[width=170mm]{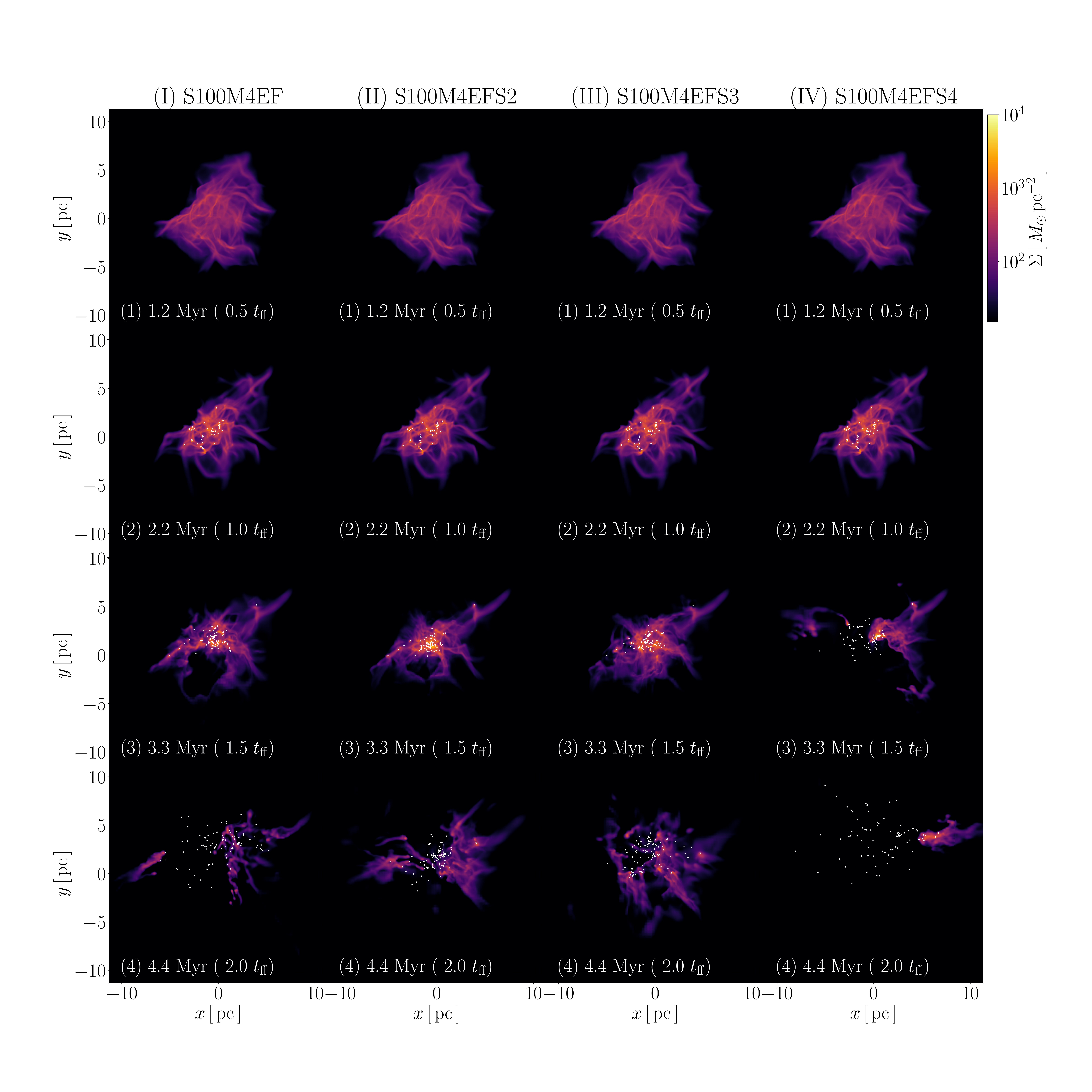}
    \end{center}
    \caption{
    The evolution of surface densities in the clouds with $(\Sigma_{\rm cl}, M_{\rm cl}) = (10^2~M_{\odot} {\rm pc^{-2}}, 10^4~M_{\odot})$.
    Each column shows the models of (I) S100M4EF, (II) S100M4EFS2, (III) S100M4EFS3, and (IV) S100M4EFS4.
    As in Figure \ref{fig_sigma100m4}, white dots represent the positions of star particles.
}
    \label{fig_sig100m4seed}
\end{figure*}

As discussed in Section \ref{Sec_rad_prop}, the total stellar mass is not high enough to fulfill the mass distribution of the IMF in the clouds with $10^{4}~M_{\odot}$.
The emissivity of the star cluster
depends on the stellar population determined by stochastic sampling using random numbers.
Therefore, the stellar populations can change depending on the initial random number seed.
To investigate the sampling effects in the low-mass clouds, we perform the additional simulations of the model S100M4EF with the different seeds.
The additional models are labeled as S100M4EFS2.
We also performs the simulations of the models with $(\Sigma_{\rm cl}, M_{\rm cl})=(10^2~M_{\odot} {\rm pc^{-2}}, 10^6~M_{\odot})$ to investigate the effects of stochastic stellar population in high-mass clouds.

Figure \ref{fig_sig100m4seed} shows the evolution of the surface densities with the different seeds.
The distributions of gas and star particles are almost the same until $t\sim t_{\rm ff}$.
After that, the expansion of H{\sc ii} regions starts, and the effects of the different stellar populations appear.
In the models of S100M4EF and S100M4EFS2, the star clusters make the cavities by the photoionization feedback at $t \sim 1.5 ~t_{\rm ff}$.
In the model of S100M4EFS4, the cloud disruption proceeds earlier.
On the other hand, in the case of S100M4EFS3, there is no cavity even at $t \sim 1.5 ~t_{\rm ff}$.
At $t\sim 2 ~t_{\rm ff}$, the gas distributions are different depending on the random number seeds. Most gas is evacuated in S100M4EFS4, while S100M4EFS3 still keeps the gas within the initial cloud radius.

\begin{figure}
    \begin{center}
    	\includegraphics[width=\columnwidth]{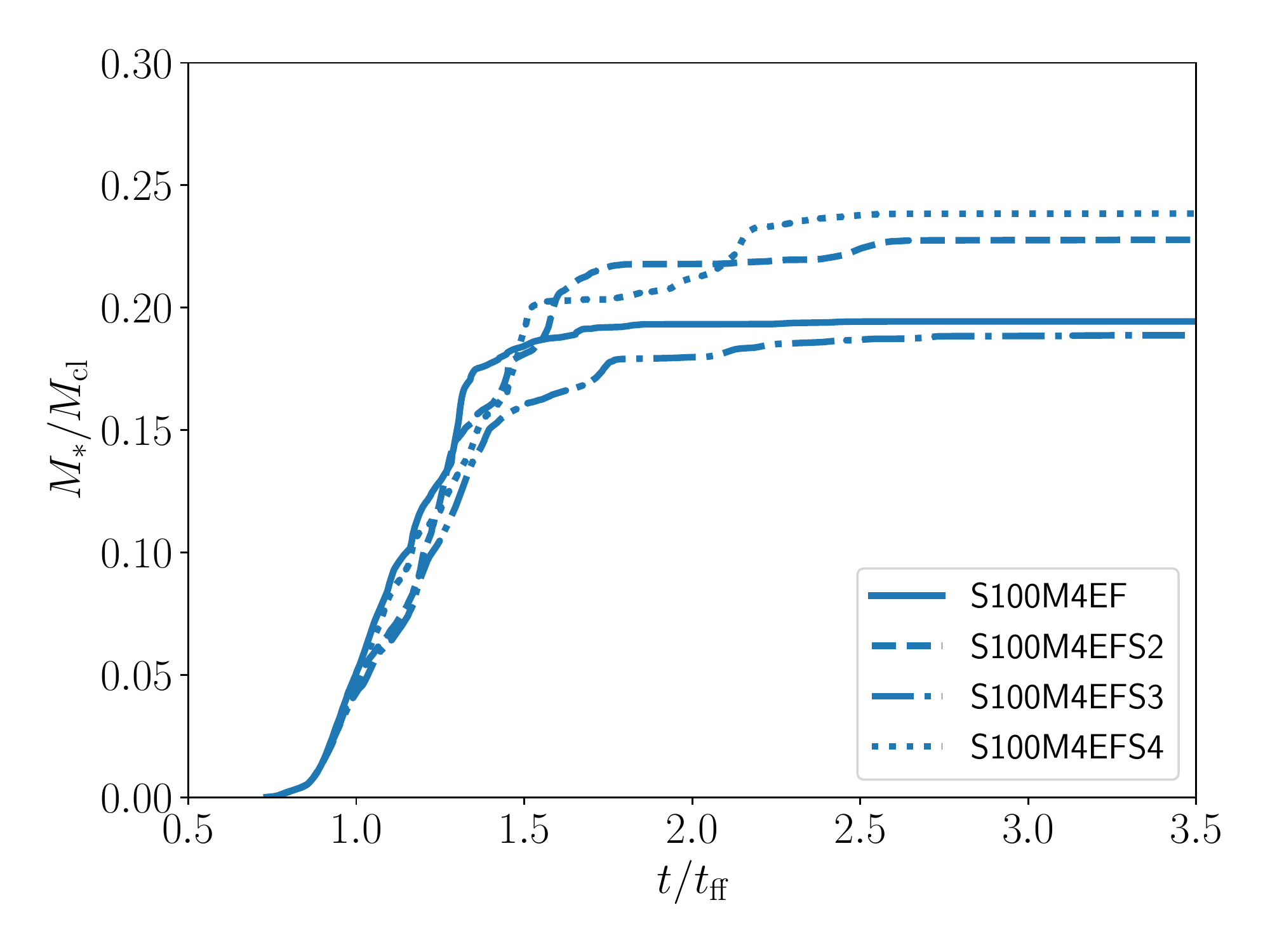}
    \end{center}
    \caption{
    The star formation histories with the different seeds in the models with $(\Sigma_{\rm cl}, M_{\rm cl})=(10^2~M_{\odot} {\rm pc^{-2}}, 10^4~M_{\odot})$.
    Each line shows the results of S100M4EF (solid), S100M4EFS2 (dashed), S100M4EFS3 (dot-dashed), and S100M4EFS4 (dot).
}
    \label{fig_mstars100m4s}
\end{figure}

The star formation histories are shown in Figure \ref{fig_mstars100m4s}.
The results are similar until $t\sim 1.3~t_{\rm ff}$.
Then, the radiative feedback affects the cloud evolution, and the total stellar mass deviates from each other.
The SFE of S100M4EFS4 is highest, in which the H{\sc ii} regions expand rapidly.
We find that the difference of the SFEs between S100M4EFS4 and S100M4EFS3 (lowest) is 5 percent.

\begin{figure}
    \begin{center}
    	\includegraphics[width=\columnwidth]{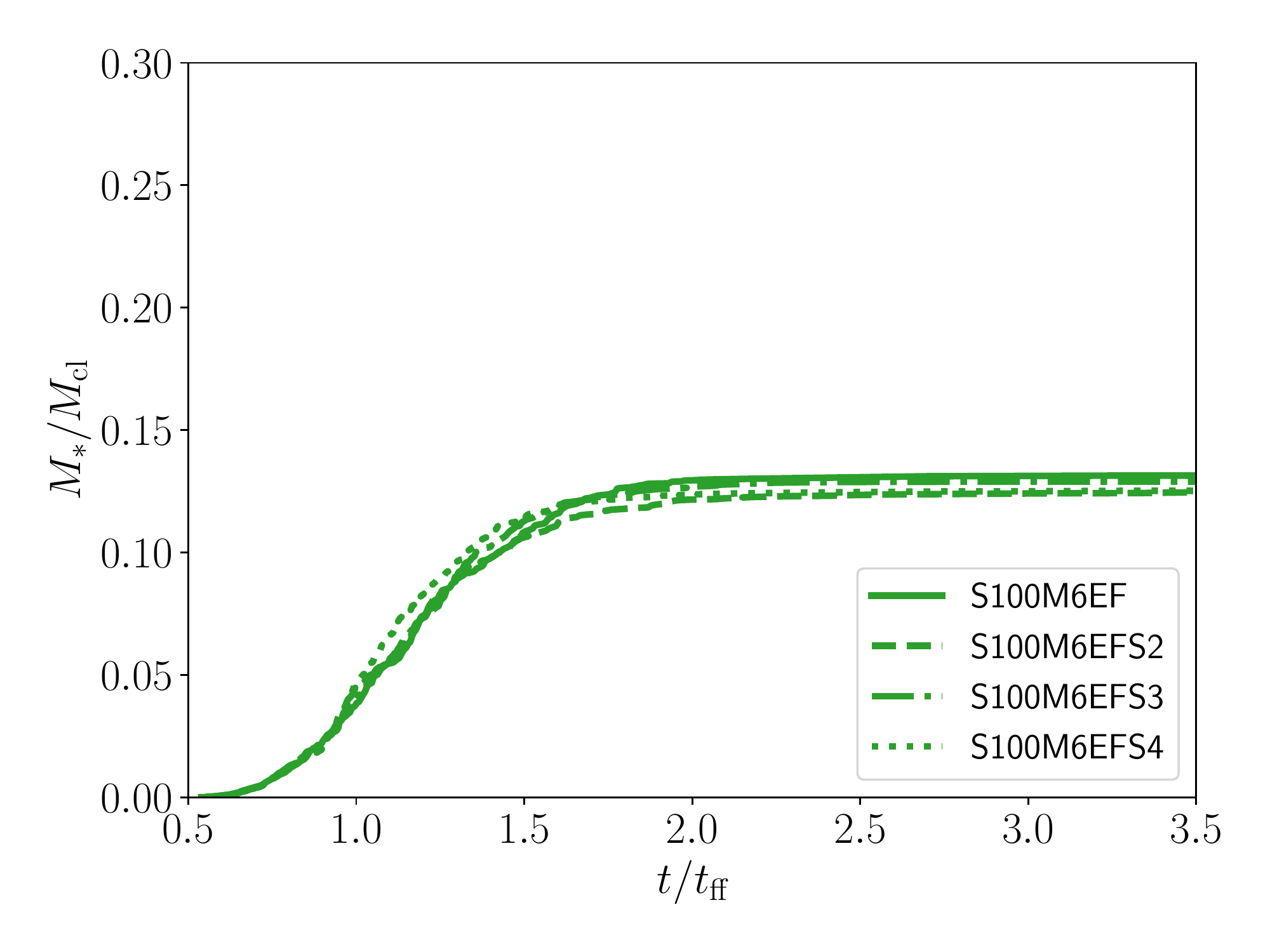}
    \end{center}
    \caption{
    Same as Figure \ref{fig_mstars100m4s}, but for the case with  $(\Sigma_{\rm cl}, M_{\rm cl})=(10^2~M_{\odot} {\rm pc^{-2}}, 10^6~M_{\odot})$.
    Each line shows the results of S100M6EF (solid), S100M6EFS2 (dashed), S100M6EFS3 (dot-dashed), and S100M6EFS4 (dot).
}
    \label{fig_mstars100m6s}
\end{figure}

To investigate the effects of the stochastic stellar population in the massive cloud with $M_{\rm cl} = 10^6~M_{\odot}$, we additionally perform the simulations of S100M6EF model with the different random seeds.
Figure \ref{fig_mstars100m6s} shows the star formation histories in each model.
We find no apparent difference in the SFE, and its value converges to $\sim 0.13$.
As discussed in Section \ref{Sec_stochastic_stellar_pop}, the stellar population of the star clusters is almost the same as the modelled IMF if the total stellar mass is larger than $10^4~M_{\odot}$.
The emissivities of star clusters in these clouds are equal to the IMF-averaged values.
We, therefore, suggest that the impacts of the stochastic stellar population can be negligible in massive clouds.

\section{mass-to-luminosity ratios}\label{app_mass_to_lumi_ratio}

As shown in Figure \ref{fig_mass_rad}, there is a large dispersion in the mass-to-luminosity ratio for the stellar masses lower than $10^4~M_{\odot}$.
In such a case, the expected number of massive stars is below unity. Therefore, the mass-to-luminosity ratio should be lower than the IMF-averaged value.
Same as \citet{2016ApJ...819..137K} and \citet{2020MNRAS.497.5061I}, we make fitting functions by using median values to the results of total FUV luminosity ($\Psi_{*, {\rm FUV}}$, $6~{\rm eV} < h\nu<13.6~{\rm eV}$) and the emissivity of EUV photons ($\Xi_{\rm EUV}$) for the analytical estimates in Section \ref{Section_analytical_arguments} as
\begin{align}
   \log  \left( \frac{\Psi_{*, {\rm FUV}}}{L_{\odot} \, M_{\odot}^{-1}} \right) = \frac{3.0 \chi^6}{2.50 + \chi^6}, \label{eq_psi_fuv}
\end{align}
\begin{align}
   \log  \left( \frac{\Xi_{\rm EUV}}{{\rm s^{-1}} \, M_{\odot}^{-1}}  \right) = \frac{46.7 \chi^6}{3.66 + \chi^6}, \label{eq_xi_euv}
\end{align}
where $\chi = \log(M_{*}/M_{\odot} )$.

\section{Thermal pressure of H{\sc ii} regions}\label{app_thermal_HII}
In \citetalias{2021MNRAS.506.5512F}, we use the expanding shell model in the analytical estimate of the SFEs.
The equation of motions of the expanding shell is given by \citep{2009ApJ...703.1352K, 2016ApJ...819..137K}
\begin{align}
    \frac{d}{dt} \left( M_{\rm sh} \dot r_{\rm sh} \right) = 4 \pi r_{\rm sh}^2 \rho_{\rm i} c_{\rm i}^2 , \label{eq_eom_of_shell}
\end{align}
where $M_{\rm sh}$ and $r_{\rm sh}$ are the mass and the radius of the shell, $\rho_{\rm i}$ and $c_{\rm i}$ are the density and and sound speed in the H{\sc ii} regions.
The total stellar mass is represented as $M_* = \epsilon_*M_{\rm sh}/(1-\epsilon_*)$ where $\epsilon_*$ is the SFE.
The first term of the right-hand side of Equation \eqref{eq_eom_of_shell} is the thermal pressure from the H{\sc ii} regions.
In the H{\sc ii} regions, the number density is determined by the balance between the ionization and recombination as
\begin{align}
    n_{\rm i} = \left( \frac{\rho_{\rm i} c_{\rm i}^2}{k_{\rm B} T_{\rm i}} \right) = \left(\frac{3 f_{\rm ion} S_{\rm ion}}{4 \pi r_{\rm sh}^3 \alpha_{\rm B}} \right)^{1/2}, \label{eq_balance_HIIregion}
\end{align}
where $T_{\rm i}$, $\alpha_{\rm B}$, and $f_{\rm ion}$ are the the temperature of ionized gas, the recombination rate coefficient, and the absorption rate of ionizing photons by hydrogen atom as $f_{\rm ion} = 0.73$ \citep{1997ApJ...476..144M, 2009ApJ...703.1352K}.
The emissivity of ionizing photons is estimated as $S_{\rm ion} =  s_{*} \epsilon_* M_{\rm sh}/(1-\epsilon_*) $.
Substituting equation \eqref{eq_balance_HIIregion} into \eqref{eq_eom_of_shell}, we obtain the force from the H{\sc ii} regions on the shell as
\begin{align}
    F_{\rm th, IF} = k_{\rm B} T_{\rm i} \left( \frac{12 \pi f_{\rm ion} \epsilon_* s_{*}  R_{\rm cl}}{\alpha_{\rm B}M_{\rm cl}} \right)^{1/2}. \label{eq_FthIF}
\end{align}
where we assume that the shell mass is comparable to the cloud mass as $M_{\rm sh} \sim M_{\rm cl}$.
The gravitational force on the shell is estimated as \citep{2016ApJ...819..137K}
\begin{align}
    F_{\rm g, sh} \sim - \frac{GM_{\rm cl}}{2 R_{\rm cl}^2}. \label{eq_Fgshell}
\end{align}
As shown in \citetalias{2021MNRAS.506.5512F}, the SFE is evaluated by assuming that the duration time of the star formation is equal to the expansion time of the H{\sc ii} regions.
Thus, we derive the SFE as
\begin{align}
    \epsilon_* &\simeq 0.09 \left( \frac{\epsilon_{\rm ff}}{0.1} \right)^{4/5} \left( \frac{\Sigma_{\rm cl}}{80~M_{\odot} {\rm pc^{-2}}}  \right)^{1/2} \left( \frac{M_{\rm cl}}{10^5~M_{\odot}} \right)^{1/10} \nonumber \\
    & \hspace{1cm} \times \left( \frac{T_{\rm i}}{8000~{\rm K}} \right)^{-14/25} \left( \frac{s_*}{7.5 \times 10^{46} ~{\rm s^{-1}M_{\odot}}} \right)^{-1/5}, \label{eq_eepps}
\end{align}
where $\epsilon_{\rm ff}$ is the parameter related with the star formation rate as $\dot M_{*} = \epsilon_{\rm ff} M_{\rm cl} /t_{\rm ff}$ where $t_{\rm ff}$ is the free-fall time.
Using the equations \eqref{eq_xi_euv}, \eqref{eq_FthIF}, \eqref{eq_Fgshell} and \eqref{eq_eepps}, we estimate the conditions that thermal pressure from H{\sc ii} regions overcomes the gravitational force at the outer edge of the cloud as shown in Figure \ref{fig_Pth_Pg}.


\bsp	
\label{lastpage}
\end{document}